\documentclass[fleqn,usenatbib]{mnras}
\usepackage{epsfig}
\usepackage{longtable}
\usepackage{color}
\usepackage{multirow}
\usepackage[T1]{fontenc}

\usepackage{graphicx}	
\usepackage{amsmath}	
\usepackage{amssymb}	

\usepackage{xcolor}
\usepackage[normalem]{ulem}

\def\gtsima{$\; \buildrel > \over \sim \;$}
\def\ltsima{$\; \buildrel < \over \sim \;$}
\def\gsim{\lower.5ex\hbox{\gtsima}}
\def\lsim{\lower.5ex\hbox{\ltsima}}
\def\kms{km s$^{-1}$}

\def\Lya{Ly$\alpha$}

\newcommand{\CIV}{\mbox{C\,{\sc iv}}}
\newcommand{\CV}{\mbox{C\,{\sc v}}}
\newcommand{\CIII}{\mbox{C\,{\sc iii}}}
\newcommand{\CII}{\mbox{C\,{\sc ii}}}

\newcommand{\OVI}{\mbox{O\,{\sc vi}}}

\newcommand{\OI}{\mbox{O\,{\sc i}}}

\newcommand{\MgII}{\mbox{Mg\,{\sc ii}}}

\newcommand{\SiIV}{\mbox{Si\,{\sc iv}}}
\newcommand{\SiIII}{\mbox{Si\,{\sc iii}}}

\newcommand{\SiV}{\mbox{Si\,{\sc v}}}

\newcommand{\FeII}{\mbox{Fe\,{\sc ii}}}

\newcommand{\HI}{\mbox{H\,{\sc i}}}

\newcommand{\HeII}{\mbox{He\,{\sc ii}}}

\title[The evolution of the \SiIV\ content in the Universe]{The evolution of the \SiIV\ content in the Universe from the epoch of reionization to cosmic noon}
\author[V. D'Odorico, et
  al.]{V. D'Odorico$^{1,2,3}$\thanks{E-mail:valentina.dodorico@inaf.it}, 
K. Finlator$^{4,5}$, S. Cristiani$^{1,3}$, G. Cupani$^{1,3}$, S. Perrotta$^6$, F. Calura$^{7}$, \and M. C\`enturion$^1$,  G. Becker$^{8}$, T. A. M. Berg$^{9,10}$, S. Lopez$^{10}$, S. Ellison$^{11}$, E. Pomante$^1$ \\ 
$^1$ INAF- Osservatorio Astronomico di Trieste, Via Tiepolo 11, I-34143 Trieste, Italy \\
$^2$ Scuola Normale Superiore di Pisa, Piazza dei Cavalieri 7, I-56126 Pisa, Italy \\
$^3$ IFPU–Institute for Fundamental Physics of the Universe, via Beirut 2, I-34151 Trieste, Italy \\
$^4 $New Mexico State University, Las Cruces, NM, USA\\
$^5$ Cosmic Dawn Center (DAWN), Niels Bohr Institute, University of Copenhagen / DTU-Space, Technical University of Denmark \\
$^6$ Department of Astronomy, University of California, San Diego, CA 92092, USA \\
$^7$ INAF - Osservatorio Astronomico di Bologna, Via Ranzani 1, 40127 Bologna, Italy
\\
$^8$ Department of Physics \& Astronomy, University of California, Riverside, CA 92521, USA \\
$^9$ ESO - European Southern Observatory, Alonso de Cordova 3107, Casilla 19001, Santiago, Chile\\
$^{10}$ Departamento de Astronom\'\i a, Universidad de Chile, Casilla 36-D, Santiago, Chile \\
$^{11}$ Department of Physics \& Astronomy, University of Victoria, Finnerty Road, Victoria, British Columbia V8P 1A1, Canada
}

\date{Accepted XXX. Received YYY; in original form ZZZ}

\pubyear{2021}

\begin{document}
\label{firstpage}
\pagerange{\pageref{firstpage}--\pageref{lastpage}} 

\maketitle

\begin{abstract}
 We investigate the abundance and distribution of metals in the high-redshift intergalactic medium and circum-galactic medium through the analysis of a sample of almost 600 \SiIV\ absorption lines detected in high and intermediate resolution spectra of 147 quasars. The evolution of the number density of \SiIV\ lines, the column density distribution function and the cosmic mass density are studied in the redshift interval $1.7 \lsim z \lsim 6.2$ and for $\log N($\SiIV$) \ge 12.5$.  All quantities show a rapid increase between $z\sim6$ and $z\lsim 5$ and then an almost constant behaviour to $z\sim2$ in very good agreement with what is already observed for \CIV\ absorption lines. The present results are challenging for numerical simulations:
when simulations reproduce our \SiIV\ results, they tend to underpredict the properties of \CIV, and  when the properties of \CIV\ are reproduced, the number of strong \SiIV\ lines ($\log N($\SiIV$) > 14$) is overpredicted.
\end{abstract}


\section{Introduction}
Galaxies evolve through continuous exchanges of gas with the surrounding circumgalactic medium (CGM) and intergalactic medium (IGM). 
In the scenario suggested by state-of-the-art  simulations, galaxies at the cosmic noon ($z \sim 2-3$) accrete a substantial fraction of their gas through “cold flows”, dense filamentary accretion streams that flow nearly unaffected through the galaxy halo and provide cold, $T \sim 10^4$ K, gas accretion to the interstellar medium \citep[ISM; e.g.][]{birnboim03,keres05,fauchergiguere11,vandevoort11,nelson13,theuns21}. 
At the same time, the vigorous star formation activity going on in these high-$z$ galaxies favours the dispersion in the CGM (and eventually, IGM) of the metal enriched gas through feedback mechanisms driven by galactic winds, SN explosions and accretion onto black holes \citep[e.g.][]{madau01,calura06,barai13,suresh15,turner16,muratov17,fossati21}. 

 While the observational evidence of inflowing cold flows is still tentative (e.g. Rubin et al. 2012), the presence of metal enriched gas outside galaxies has been well established observationally for more than two decades \citep[e.g.][]{cowie95,tytler95,ellison00,schaye03,dodorico16}.   

The physical properties and chemical composition of the diffuse gas are probed by absorption-line spectroscopy of bright background sources. In particular,  metal absorption lines associated with \HI\ Lyman-$\alpha$ (Ly$\alpha$) clouds with column densities $\log N($\HI$) \lsim 17.3$    are thought to arise in gas located on the outskirts of galaxies and further away, in the shallow IGM, as the \HI\ column density decreases~\citep{vandevoort12}.

Several studies have investigated the properties of metals close to galaxies by considering the
correlation between the metal absorptions observed along a quasar line of
sight and the galaxies present in the field at matching redshifts. All
the studies carried out up to now both at high \citep[e.g.][]{adelberger,steidel10,turner14,lofthouse19,rudie19} and low
redshift \citep[e.g.][]{tumlinson11,prochaska11,werk14,liang14,bordoloi14,johnson17,fossati19} agree on the significant presence of metals in high and low ionization state
at impact parameters at least as large as $\approx 100-300$ kpc. 
The standard way to carry out this kind of investigation has been by considering samples of galaxy-absorber pairs and deducing the metal distribution in the CGM from a statistical point of view. With the advent of the 30-40m class telescopes (e.g. the ESO Extremely Large Telescope\footnote{https://elt.eso.org}), it will be possible to use much fainter sources as background targets for spectroscopy (in particular, galaxies themselves) and pierce the gas surrounding the same galaxy with multiple, close lines of sight  to carry out a tomographic study of the CGM. The few examples  available today reveal the power of this technique to assess the gas patchiness and its covering factor around individual systems  \citep[e.g.][]{lopez18,lopez19}. 

The distribution of metals and the enrichment mechanism at high redshift can be investigated also with large samples of absorption lines detected along many, independent lines of sight to bright sources. The statistical properties derived from those samples can then be compared with model predictions to constrain in particular the adopted feedback mechanisms \citep[e.g.][]{tescari11}. 
The most studied ionic transition at high redshift is the triply ionized carbon doublet  (\CIV\ $\lambda\lambda\ 1548, 1551$ \AA) mainly due to the fact that: {\it  i)} it is commonly observed; {\it ii)} it falls outside the \Lya\ forest; {\it iii)} it is easy to identify thanks to its doublet nature; {\it iv)} it is observable with ground-based telescopes from $z\simeq1.0$ to the highest redshift at which we can observe quasars to date.
\CIV\ traces ionized gas mostly in the CGM and IGM environment \citep[see e.g.][]{opp_dave06,cen_chisari11,mongardi18}.   
The \CIV\  number density, column density distribution function and cosmic mass density, $\Omega_{\rm CIV}$,  are the statistical indicators mostly used to assess the evolution of the abundance of this ion across the history of the Universe 
\citep[e.g.][]{scannapieco,ryanweber09, cooksey10, dodorico10, simcoe11, cooksey13, dodorico13, codoreanu18, meyer19,  cooper19,hasan20}.  
Cosmological hydro-dynamical simulations by several groups have tried to reproduce the measured observables (mainly the column density distribution function and the cosmic mass density) all with poor results, in particular for $z \gsim 5$ \citep[e.g.][]{rahmati16, keating16}. The introduction of a fluctuating ultra-violet background (UVB) improves the agreement but does not solve the problem \citep[][and references therein]{finlator18}. On the other hand, \citet{garcia17} are able to reproduce the column density distribution function of \CIV\ at $z \geq 4.35$, but at the price of an overproduction of \SiIV. This overabundance of high column density \SiIV\ absorbers is not present in \citet{rahmati16} and \citet{finlator18}. 

There are clear numerical limitations in the realization of the optimal \CIV\ simulation: it should both reproduce enough strong absorbers (need of a large box) and a self consistent, inhomogeneous UVB, implementing radiative transfer (computationally very expensive). In addition to the uncertainties in stellar yields and feedback mechanisms, the amount of \CIV\ is made even more difficult to determine by the fact that the ionization potentials to convert \CIII\ into \CIV\ and  \CIV\ into \CV\ are on the opposite sides of the UVB jump  due to the ionization of \HeII, at 4 Ryd or 54.4 eV. The shape of the UVB at energies $> 4$ Ryd is very uncertain \citep[see e.g. Fig.~9 in][]{finlator18}. 

Another tracer of the ionized diffuse gas in the CGM is the triply ionized silicon  \citep[\SiIV\ $\lambda\lambda\ 1394, 1403$ \AA, see e.g.][]{steidel10, turner14, mongardi18}, which is not as common as \CIV\ but it shares with the latter the advantage of being a doublet that is observable outside the \Lya\ forest. \SiIV\ is probably tracing slightly denser gas with respect to \CIV, and thus possibly closer to galaxies \citep[e.g.][]{mongardi18}. However, among the absorption lines routinely detected in quasar spectra (e.g. \CIV, \SiIV, \OVI, \MgII, \FeII) it is the one that arises from gas in a range of densities and temperatures almost equivalent to those of gas traced by \CIV\ \citep[see e.g. Fig. 12 in][]{scannapieco}. \CIV\ and \SiIV\ are the dominant ionization stages of carbon and silicon in the IGM, and furthermore they are tracers of Fe-coproduction and $\alpha$-element processes, respectively. \SiIV\ can be observed from the ground for $z\gsim 1.15$.  A bonus of \SiIV\ with respect to \CIV\ is that the ionization potentials to turn \SiIII\ into \SiIV\ and \SiIV\ into \SiV\ fall in the energy range $2 < IP < 4$ Ryd, where the different UVB models  behave very similarly.          

In her pioneering work, \citet{songaila2001} analysed both \CIV\ and \SiIV\ absorption lines for a sample of 32 quasars observed with HIRES and ESI at the Keck telescope, covering the redshift range $1.7 \lsim z \lsim 5.3$. Considering \SiIV\ lines with column density $12.00 \le \log N($\SiIV$) \le 14.8$, she found that the cosmic mass density of this ion, $\Omega_{\rm SiIV}$, keeps approximately constant at lower redshift with a downturn at $z > 4.5$ (see Fig.~\ref{fig_omega}). Subsequent works that investigated the properties of \SiIV\ absorbers were all based on smaller samples and/or narrower redshift ranges \citep[e.g.][]{scannapieco, cooksey11, shull14, BS15, codoreanu18}. In general, they all confirmed the redshift behaviour of $\Omega_{\rm SiIV}$ observed by \citet{songaila2001}.  Observationally, \SiIV\ broadly matches the behaviour of \CIV\ in terms of redshift evolution of the cosmic mass density and line number density \citep[e.g.][]{songaila2001, BS15}. Also the clustering properties of \SiIV\ absorption lines along the line of sight are in very good agreement with those of \CIV\  suggesting similar sizes for the ion enriched bubbles \citep{scannapieco}. 

For all the reasons explained above and because of the large number of available quasar lines of sight observed at intermediate to high resolution and good signal-to-noise ratio (SNR), we believe it is timely to dedicate a paper to \SiIV\ and its statistical properties.  In this work, we present a sample of 519 \SiIV\ absorption lines with column density $\log N$(\SiIV$) \ge 12.5$ observed along 147 quasar lines of sight and covering a very broad redshift range, $z\simeq 1.75-6.24$. 
Our aim is to study in detail the evolution of this ion, contrasted it with \CIV\ and the recently studied \OI\ \citep{becker19}, and provide solid observational results to  be compared with simulation predictions to help figuring out the physics of the feedback mechanisms.
The detailed study of the \SiIV\ and \CIV\ absorptions arising in the same systems (e.g. the determination and analysis of the redshift evolution of the \SiIV/\CIV\ ratio) will be deferred to a future work. 

The paper is organized as follows. Section 2 describes the three spectroscopic samples that have been used in this work.  In Section 3, we report the results for the line number density and column density distribution function computed with our sample. Section 4 is devoted to the presentation of the results on the cosmic mass density and its evolution with redshift. In section 5, we compare our results with previous works appeared in the literature. Finally, Section 6 is dedicated to the discussion and conclusions.  
Throughout this paper, we assume $\Omega_{\rm m} = 0.3$,
$\Omega_{\Lambda} = 0.7$ and $h \equiv H_0/(100 {\rm km\ s}^{-1} {\rm
  Mpc}^{-1}) =0.70$ if not stated otherwise.           

\begin{figure}
\begin{center}
\includegraphics[width=9cm]{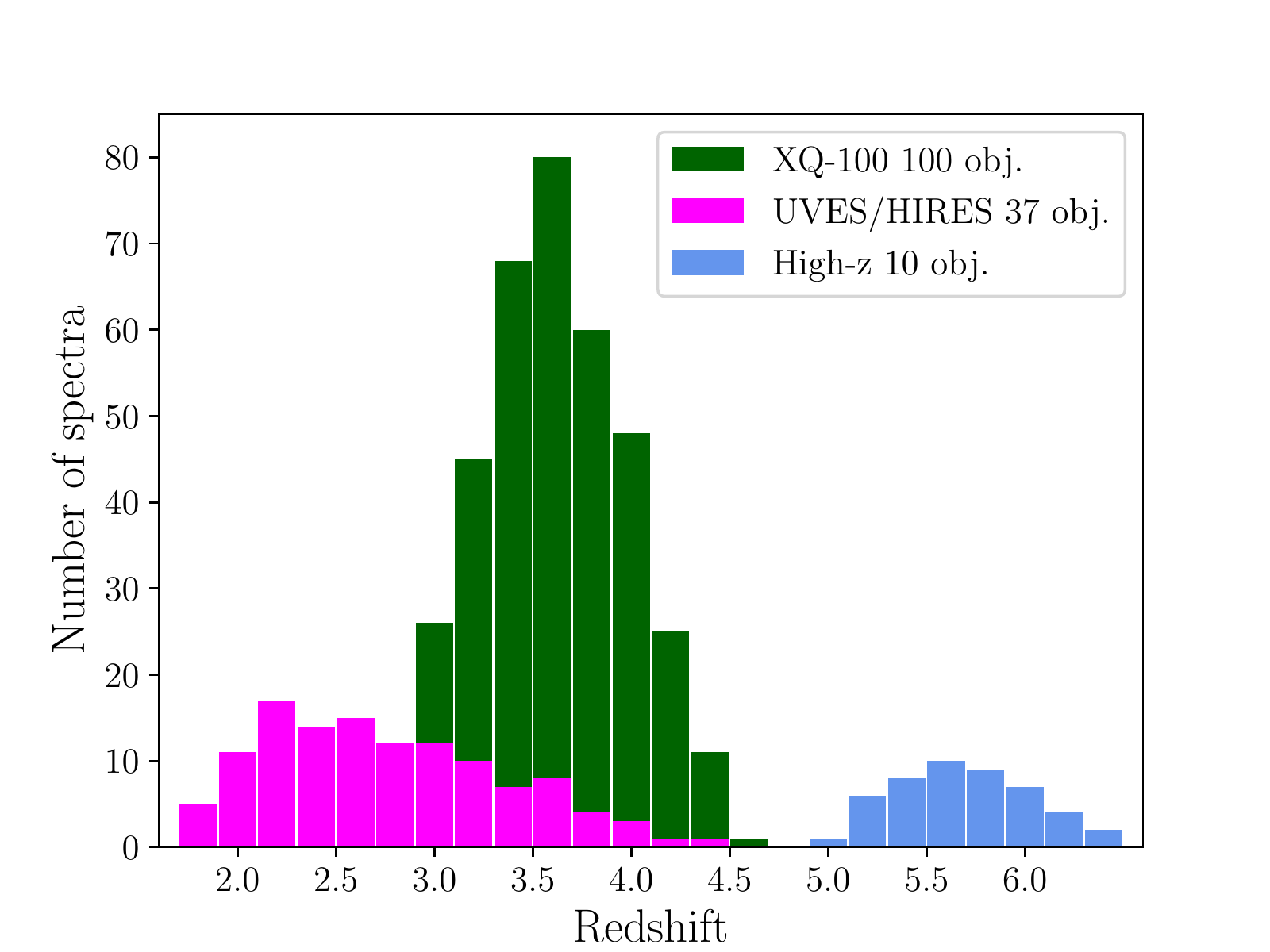}
\caption{Number of spectra used in this work per redshift bin of
  $\Delta z = 0.2$. }
\label{fig_nspec}
\end{center}
\end{figure}

\section{Data samples and analysis}
This work is based on three samples of quasar spectra, for a total of 147 lines of sight, which cover
different redshift ranges for \SiIV\ absorptions. They are briefly  described in the following sections.
 
In all the considered spectra, \SiIV\ doublets were looked for by eye in the region outside the \Lya\ forest, redward of the quasar
\Lya\ emission. In case of doubtful detections (e.g. blending of one
of the two components of the doublet), we used the presence of a
\CIV\ system at the same redshift and with a similar velocity profile
to confirm the identification.  To avoid contamination from absorption systems
associated with the quasars, we considered only absorbers
with a maximum redshift, $z_{\rm max}$, at a separation $\Delta v =
-5000$ \kms\ from the quasar emission redshift. On the other hand, the
contamination from \Lya\ lines was prevented by taking a minimum
redshift, $z_{\rm min}$, at a separation $\Delta v = +1000$ \kms\ from
the \Lya\ emission of the quasar.  
The considered redshift intervals are reported in
Tabs.~\ref{tab_highres}, \ref{tab_XQ100} and \ref{tab_highz}.  

All the detected \SiIV\ absorption lines were fitted with Voigt profiles
mainly with the {\sl fitlyman} context of the ESO {\small MIDAS}
package \citep{font:ball}.  Exceptions are discussed in the  sections dedicated to the sample description.  

The number of spectra per redshift bin of $\Delta z = 0.2$ for the
three samples, is shown in Fig. \ref{fig_nspec}. 

\subsection{The high-resolution UVES/HIRES sample}
\label{sec_highres}
This sample collects the high-resolution ($R\simeq 50,000$, $\Delta v \simeq 6.0$ km s$^{-1}$) spectra analysed in \citet{dodorico10} and \citet{calura12}. For the quasar HE 0940-1050, we used the higher SNR spectrum presented in \citet{dodorico16}. Details of the data reduction and analysis are reported in \citet{dodorico10}. 

Also, we used updated spectra (including new observations) of SDSS J1621-0042 and PKS 1937-101 and added the new spectra of QSO J0407-4410 and  QSO J0103+1316.  
The four new spectra  were reduced by E. Pomante in the context of his PhD thesis work\footnote{http://hdl.handle.net/11368/2908079}, with a custom made pipeline able to ingest both UVES and HIRES raw frames and treat them in the same way. This generalized pipeline is an extension of the IDL code developed by S. Burles and J. X. Prochaska for the reduction of MIKE and HIRES data \citep{Bernstein15}, with a modified approach 
to wavelength calibration, flat fielding and object extraction.

In order to increase the statistical significance of our sample and the coverage toward higher redshifts, we have considered also the sample of 9 objects analysed in \citet[][hereafter BS15]{BS15}, which have a similar resolution element of $\Delta v \simeq 6.6$ km s$^{-1}$. For the quasar spectrum of B1422+231, which is in common between our sample and BS15, we considered our list of lines, which is consistent with the BS15 one. 

Note that the majority of the quasars in this sample belongs to the UVES Large Programme by \citet{bergeron04} which required Lyman-$\alpha$ forests free from Damped Lyman-$\alpha$ systems (DLAs). We have thus verified the presence of DLAs in the additional quasars and excised from the final sample the redshift ranges of the corresponding \SiIV\ lines. In particular, there is a DLA per line of sight in the BS15 spectra of QSO B1425+6039, QSO B1055+4611 and QSO B2237-0607.    

The final high-resolution sample is then formed by 37 objects and covers the \SiIV\ redshift range  $1.75 \lsim z \lsim 4.47$, with a total scanned absorption path of $\Delta X \simeq 54$. In Table~\ref{tab_highres}, we report for each object of the sample the redshift range available for \SiIV\ detection and the SNR per resolution element of 6.6 km s$^{-1}$ at $\lambda_{\rm rest} \sim 1380$ \AA, to be compliant with the SNR reported by BS15. The average SNR~$\sim156$ corresponds to a $3\,\sigma$ detection threshold for the \SiIV\ $\lambda 1393$ \AA\ line of $\log N($\SiIV$)_{\rm thr} \simeq 11.09$ assuming a conservative $b =10$ \kms\ Doppler parameter. The spectrum with the lowest SNR has $\log N($\SiIV$)_{\rm thr} \simeq 11.95$.

BS15 fit absorption lines with the  Voigt profile fitting code {\small VPFIT} \citep{vpfit}. As already shown in
previous works \citep[e.g.][]{dodorico16} there are no significant differences between the results of {\sl fitlyman} and {\small VPFIT}.
Possible differences due to, for example, an excess of low column density components by {\small VPFIT}, are further mitigated by the fact that, as we already did in \citet{dodorico10,dodorico13}, in order to compare these high-resolution 
spectra with those from XSHOOTER, \SiIV\ lines with velocity separation smaller than 50 km s$^{-1}$ have been  merged. 
The merging process has been carried out in the following way: for each list of \SiIV\ components corresponding to a single quasar, the velocity separations among all the lines have been computed and sorted in ascending order. If the smallest separation is less than $dv_{\rm min} = 50$ km s$^{-1}$, the two corresponding absorption lines are merged into a new line with column density equal to the sum of the column densities, and redshift equal to the average of the redshifts, weighted by the column density of the components. The velocity separations are then computed again and the procedure is iterated until the smallest separation becomes larger than $dv_{\rm min}$.
For consistency, the same merging process was applied to the XQ-100 sample and to the $z\sim6$ XSHOOTER sample described in Sec.\ref{sec_XQ100} and \ref{sec_highz}, respectively.   


\begin{table}
\caption{The high-resolution UVES/HIRES sample. The different columns report: the name of the object, the emission redshift, the minimum and maximum redshift considered for the \SiIV\ absorption lines to be part of the sample, the signal-to-noise ratio per resolution element computed at $\lambda_{\rm rest} \sim 1380$ \AA\ and the reference paper. }
\begin{minipage}{75mm}
\label{tab_highres}
\begin{tabular}{l l l l r c}
\hline  
Object & $z_{\rm em}$ & $z_{\rm min}$ & $z_{\rm max}$ &  SNR & Ref. \\ 
\hline
HE 1341-1020 &  2.142 &  1.749  & 2.090  & 200 &1 \\
QSO B0122-379    &  2.200 &  1.801  & 2.147  & 114 & 1 \\
PKS 1448-232 &  2.224 &  1.821  & 2.171  & 113 & 1 \\
PKS 0237-230  &  2.233 &  1.829  & 2.179  &  283 &1 \\
HE 0001-2340 &  2.267 &  1.859  & 2.213  & 190 & 1 \\
QSO B1626+6426 & 2.320 &  1.905 & 2.265 & 128 & 5 \\
HE 1122-1648 &  2.400 &  1.975  & 2.344  &  314 & 1 \\
QSO B0109-3518  &  2.406 &  1.980  & 2.349  &  119 & 1 \\
HE 2217-2818 &  2.414 &  1.988  & 2.357  &  246 &1 \\
QSO B0329-385   &  2.437 &  2.008  & 2.380  &  75 &1 \\
HE 1158-1843 &  2.448 &  2.017  & 2.391  &  97 &1 \\
HE 1347-2457 &  2.599 &  2.149  & 2.539  &  313 & 1\\
QSO B1442+2931 & 2.660 & 2.203 & 2.599 & 107 & 5 \\
QSO B0453-423  &  2.669 &  2.211  & 2.608  &  194 & 1 \\
PKS 0329-255 &  2.696 &  2.234  & 2.635  &  97 &1 \\
QSO J0103+1316 & 2.721 & 2.256 & 2.659 & 218 & 4 \\
HE 0151-4326 &  2.763 &  2.293  & 2.701  &  182 &1\\
QSO B0002-422   &  2.769 &  2.298  & 2.707  &  211 & 1 \\
HE 2347-4342 &  2.880 &  2.395  & 2.816  &  164 & 1 \\
QSO B1107+4847 & 2.970 & 2.474 & 2.904 & 94 & 5 \\
QSO J0407-4410 & 3.021 & 2.519 &  2.954  & 235 & 4 \\
HS 1946+7658 &  3.058 &  2.551  & 2.991 & 156 & 1 \\
HE 0940-1050 &  3.093 &  2.582  & 3.025  &  333 & 1,2\\
QSO B0420-388  &  3.126 &  2.610  & 3.057 & 138 &1\\
QSO B0636+6801 & 3.180 & 2.658 & 3.111 & 107 & 5 \\
QSO B1425+6039 & 3.180 & 2.658 &  3.111 & 140 & 5 \\ 
QSO B1209+0919  &  3.291 &  2.755  & 3.220 & 30 & 3\\
PKS 2126-158 &  3.292 &  2.756  & 3.221  & 198 &1\\
QSO B1422+2309   &  3.623 &  3.046  & 3.546 & 90  &1,5 \\
SDSS J1249-0159  &  3.630 &  3.052  & 3.553  & 85 &3\\
QSO B0055-269   &  3.660 &  3.078  & 3.583   & 142 & 1\\
SDSS J1621-0042 & 3.710 & 3.122 & 3.632 & 159 &3,4\\
QSO J1320-0523  & 3.717  & 3.128  & 3.639   & 154 & 3\\
PKS 1937-101 & 3.787  & 3.189  & 3.708  & 138 & 1,4\\
QSO J1646+5514 & 4.100 & 3.463 & 4.016 & 119 & 5 \\ 
QSO B1055+4611 & 4.15 & 3.507 & 4.065 & 47 & 5 \\ 
QSO B2237-0608 & 4.56  & 3.866 & 4.468  & 42 & 5 \\ 
\hline
\end{tabular}
1 \citet{dodorico10}; 2 \citet{dodorico16}; 3 \citet{calura12}; 4 This work; 5 \citet{BS15}
\end{minipage}
\end{table}

\begin{center}
\begin{table*}
\caption{The XQ-100 quasar sample. The columns are the same as in Table~\ref{tab_highres} with the exception of the reference column, since all the spectra were presented in \citet{lopez16} and the column $R_{\rm new}$ which reports the adopted resolving power if different from the nominal one.}
\begin{minipage}{150mm}
\label{tab_XQ100}
\begin{tabular}{l l l l c r l l l l r}
\hline  
Object & $z_{\rm em}$ & $z_{\rm min}$ & $z_{\rm max}$ & $R_{\rm new}$ & & Object & $z_{\rm em}$ & $z_{\rm min}$ & $z_{\rm max}$ & $R_{\rm new}$\\ 
\hline
J1332+0052 & 3.5082  &   2.9453  &  3.4336 &   $-$   & & J0211+1107 & 3.9734  &   3.3524  &  3.8911 & 10700 \\
J1018+0548 & 3.5154  &   2.9516  &  3.4407 &  10400  & & J0214-0518 & 3.9770  &   3.3556  &  3.8947 & 9400  \\
J1201+1206 & 3.5218  &   2.9572  &  3.4470 &  13400 & & J1032+0927 & 3.9854  &   3.3629  &  3.9029 & 10300 \\
J1024+1819 & 3.5243  &   2.9594  &  3.4495 &  9800 & &  J1542+0955 & 3.9863  &   3.3637  &  3.9038 & 10400 \\
J1442+0920 & 3.5319  &   2.9660  &  3.4569 &  9900 & & J2215-1611 & 3.9946  &   3.3710  &  3.9120 & 15000 \\
J0100-2708 & 3.5459  &   2.9783  &  3.4707 &   9900 & & J0255+0048 & 4.0033  &   3.3786  &  3.9205 &10400 \\
J1517+0511 & 3.5549  &   2.9862  &  3.4796 &  $-$  & & J0835+0650 & 4.0069  &   3.3817  &  3.9241 & 11600 \\
J1445+0958 & 3.5623  &   2.9926  &  3.4868 &  10300 & & J0311-1722 & 4.0338  &   3.4053  &  3.9505 & 9700 \\
J1202-0054 & 3.5924  &   3.0190  &  3.5164 &    $-$ &  &  J1323+1405 & 4.0537  &   3.4227  &  3.9701 & 10100 \\
J1416+1811 & 3.5928  &   3.0193  &  3.5168 &   $-$ &&  J0244-0134 & 4.0546  &   3.4235  &  3.9710 & 12000 \\
J1524+2123 & 3.6002  &   3.0258  &  3.5241 &   $-$ &  &   J0415-4357 & 4.0732  &   3.4397  &  3.9893 & 9900  \\
J1103+1004 & 3.6070  &   3.0318  &  3.5308 &   $-$ &   &  J0048-2442 & 4.0827  &   3.4481  &  3.9986 & 11400 \\
J1117+1311 & 3.6218  &   3.0447  &  3.5453 &   $-$ & &     J0959+1312 & 4.0916  &   3.4558  &  4.0074  & $-$ \\
J1037+2135 & 3.6260  &   3.0484  &  3.5495 &   $-$ &  &   J0121+0347 & 4.1252  &   3.4852  &  4.0404 & $-$ \\
J1249-0159$^{a}$ & 3.6289  &   3.0509  &  3.5523 & $-$ && J0003-2603 & 4.1254  &   3.4854  &  4.0406 & $-$ \\
J1042+1957 & 3.6302  &   3.0521  &  3.5536 &  $-$ & &     J1037+0704 & 4.1271  &   3.4869  &  4.0423 & $-$ \\
J1126-0126 & 3.6346  &   3.0559  &  3.5579 &   $-$ & &    J1057+1910 & 4.1284  &   3.4881  &  4.0436 & 11800 \\
J0056-2808 & 3.6347  &   3.0560  &  3.5580 &   $-$ &  &   J0747+2739 & 4.1334  &   3.4924  &  4.0485 & $-$\\
J1020+0922 & 3.6401  &   3.0607  &  3.5633 &  9900 &   &   J1110+0244 & 4.1456  &   3.5031  &  4.0605 & 9700 \\
J0920+0725 & 3.6465  &   3.0663  &  3.5696 &  10100 & & J0132+1341 & 4.1523  &   3.5090  &  4.0671 & 9700 \\
J1304+0239 & 3.6481  &   3.0677  &  3.5712 &   $-$ & & J2251-1227 & 4.1575  &   3.5135  &  4.0722  & 15000 \\
J0818+0958 & 3.6564  &   3.0750  &  3.5794 &  11400 & &    J0529-3552 & 4.1717  &   3.5259  &  4.0862 & 10700 \\
J0057-2643$^{a}$ & 3.6608  &   3.0788  &  3.5837 & 9700 && J0030-5129 & 4.1729  &   3.5270  &  4.0873 & $-$\\
J0755+1345 & 3.6629  &   3.0807  &  3.5858 &   $-$ && J0133+0400 & 4.1849  &   3.5375  &  4.0991 & 12000 \\
J1053+0103 & 3.6634  &   3.0811  &  3.5863 &   $-$ &  &   J0153-0011 & 4.1948  &   3.5462  &  4.1089 & $-$ \\
J1108+1209 & 3.6789  &   3.0947  &  3.6015 &   9500 &  &   J2349-3712 & 4.2192  &   3.5675  &  4.1329 & 9700 \\
J1421-0643 & 3.6885  &   3.1031  &  3.6109 &    $-$ &   & J0403-1703 & 4.2267  &   3.5741  &  4.1402  & 12900 \\
J1503+0419 & 3.6919  &   3.1061  &  3.6143 &   $-$ & & J0839+0318 & 4.2298  &   3.5768  &  4.1433 & 9500 \\
J0937+0828 & 3.7035  &   3.1162  &  3.6257 &   10200 & &    J0247-0556 & 4.2335  &   3.5800  &  4.1469 & $-$ \\
J1352+1303 & 3.7065  &   3.1188  &  3.6286 &    $-$ &  & J0117+1552 & 4.2428  &   3.5882  &  4.1561 & $-$\\
J1621-0042$^{a}$ & 3.7112  &   3.1229  &  3.6333 & $-$ && J2344+0342 & 4.2484  &   3.5931  &  4.1616 & 10500 \\
J0833+0959 & 3.7162  &   3.1273  &  3.6382 &  9300 & & J1034+1102 & 4.2691  &   3.6112  &  4.1819 & 9100 \\
J1320-0523$^{a}$ & 3.7172  &   3.1282  &  3.6392 & 9100 && J0034+1639 & 4.2924  &   3.6316  &  4.2049 & $-$\\
J1248+1304 & 3.7210  &   3.1315  &  3.6429 &  $-$ & & J0234-1806 & 4.3046  &   3.6423  &  4.2169 & $-$ \\
J1552+1005 & 3.7216  &   3.1320  &  3.6435 &  10400 & & J0113-2803 & 4.3145  &   3.6509  &  4.2266 & 11500 \\
J1312+0841 & 3.7311  &   3.1404  &  3.6528 &  10400 & & J0426-2202 & 4.3289  &   3.6635  &  4.2408 & $-$ \\
J0935+0022 & 3.7473  &   3.1545  &  3.6688 &  $-$ & & J1058+1245 & 4.3413  &   3.6744  &  4.2529 & $-$ \\
J1658-0739 & 3.7496  &   3.1565  &  3.6710 &   $-$ & & J1633+1411 & 4.3650  &   3.6951  &  4.2763 & 10400 \\
J1126-0124 & 3.7650  &   3.1700  &  3.6862 &   $-$ && J0525-3343 & 4.3851  &   3.7127  &  4.2960 & 12800 \\
J1336+0243 & 3.8009  &   3.2014  &  3.7215 &  11400 &  & J1401+0244 & 4.4078  &   3.7326  &  4.3183 & 12200 \\
J1013+0650 & 3.8086  &   3.2082  &  3.7291 &  14400  &  &  J0529-3526 & 4.4183  &   3.7418  &  4.3287  & 9500 \\
J1135+0842 & 3.8342  &   3.2306  &  3.7542 &  $-$ & & J0955-0130 & 4.4185  &   3.7419  &  4.3289 & 10150 \\
J0124+0044 & 3.8368  &   3.2329  &  3.7568 &  $-$ &  &   J0248+1802 & 4.4390  &   3.7599  &  4.3490 & 12000 \\
J1331+1015 & 3.8522  &   3.2463  &  3.7719 &  11400 &  & J0006-6208 & 4.4400  &   3.7607  &  4.3500 & $-$ \\
J0042-1020 & 3.8629  &   3.2557  &  3.7825 &   11600 &  &  J0714-6455 & 4.4645  &   3.7822  &  4.3741 & 15000 \\
J1111-0804 & 3.9225  &   3.3079  &  3.8411 &   13400 &   & J2216-6714 & 4.4793  &   3.7951  &  4.3887 & 9700 \\
J0800+1920 & 3.9481  &   3.3303  &  3.8663 &   $-$ & & J1723+2243 & 4.5310  &   3.8404  &  4.4395  & $-$ \\
J1330-2522 & 3.9485  &   3.3306  &  3.8666 &    11300 & &   J1036-0343 & 4.5311  &   3.8405  &  4.4396 & 10300 \\
J0137-4224 & 3.9709  &   3.3502  &  3.8887 &    $-$ &   & J2239-0552 & 4.5566  &   3.8628  &  4.4647 & $-$ \\
J1054+0215 & 3.9709  &   3.3502  &  3.8887 &   11300 &   & J0307-4945 & 4.78    &   4.0583  &  4.6844 & 10300 \\

\hline
\end{tabular}
\end{minipage}
\end{table*}
\end{center}


\subsection{The XQ-100 sample} 
\label{sec_XQ100}

XQ-100 is a collection of 100 XSHOOTER spectra of quasars with $z_{\rm em} \simeq 3.5 - 4.5$ observed in the context 
of the ESO Large Programme ``Quasars and their absorption lines: a legacy survey of the high-redshift Universe with VLT/X-shooter'' (P.I. Sebastian L\'opez). The list of objects is reported in Table~\ref{tab_XQ100}. The sample spans the \SiIV\ redshift interval $2.94 \le z \le 4.70$. 

Spectra were obtained with a binning $\times2$ along the dispersion direction and a slit of 0.9 arcsec in the VIS  arm (where all the \SiIV\ lines fall) corresponding to a nominal resolving power of $R\sim8900$. 
Data reduction was carried out with a custom pipeline, while the manually placed continuum was determined by
selecting points along the quasar continuum free of absorption (by eye) as knots for a cubic spline. Ninety percent of the spectra has SNR~$\ge 20$ per pixel of $11$ \kms\ at 1700 \AA\ rest frame. All the details can be found in \citet{lopez16}.

Our \SiIV\ line sample was built based on \citet{perrotta16} and \citet{berg21}. We revised those detections and fitted them with Voigt profiles, using the  {\sl fitlyman} context of the ESO {\small MIDAS} package \citep{font:ball}.
During the line fitting process, we realized that many spectra had a resolving power significantly larger than the nominal one. As a consequence, we recomputed the value of $R$ for each spectrum based on the average value of the seeing during the observations (reported in the ESO archive as $DIMM$) and assuming a linear relation between resolving power and slit width (equivalent in this case to the seeing average value). If $\langle DIMM \rangle < 0.9$ arcsec, the new resolving power is obtained as $R_{\rm new} \sim (0.9 / \langle DIMM \rangle) \times 8800$. The new resolving powers adopted in the fit are reported in Table~\ref{tab_XQ100}.  The resolution element varies between $\sim20$ and 34 \kms.  

 For consistency with the high-resolution sample, we excluded from the XQ-100 \SiIV\ collection those systems which are associated with the known DLAs studied in \citet{berg16}. Our fiducial sample covers an absorption path of $\Delta X \simeq 221.8$ for a total of 385 detected \SiIV\ absorption lines with $\log N($\SiIV)$ \ge 12.5$. In the following, this sample will be called "XQ-100-noDLA". We verified that the results are not significantly affected by the inclusion or exclusion of the \SiIV\ lines associated with the DLAs.  

\subsection{The $z\sim 6$ XSHOOTER sample} 
\label{sec_highz}

This is a collection of ten XSHOOTER quasar spectra of which: 6 were already analysed in \citet{dodorico13}. ULAS J0148+0600 was analysed in \citet{codoreanu18} but we re-analysed the XSHOOTER spectrum in the ESO archive, and the last 3 spectra are published in this paper for the first time. The sample is reported in Table \ref{tab_highz} and the log of observation for the new objects is in Table~\ref{tab_obslog}. In this sample, the  \SiIV\ is observable in the redshift range $4.96 \le z \le 6.40$.   As for the XQ-100 sample, we have revised the resolution of the spectra based on the atmospheric conditions in which they were observed. The new resolving powers are reported in Table  \ref{tab_highz}.

The three new quasars were reduced with the custom pipeline used for the XQ-100 spectra and analysed with the python-based software {\sl astrocook}\footnote{http://github.com/DAS-OATs/astrocook} \citep{cupani20}. We carried out line detection, continuum determination and then line identification and Voigt profile fitting in the context of {\sl astrocook}. For consistency, we used {\sl astrocook} also to repeat the fit of the absorption lines in the other quasars of the sample. The line parameters obtained with {\sl astrocook} are consistent within the measurement errors with those reported in \citet{dodorico13} that were obtained within the {\sl fitlyman} context of the ESO {\small MIDAS}
package \citep{font:ball}.  
The 10 spectra allow to probe \SiIV\ absorption lines with $\log N($\SiIV)$ \ge 12.5$ along an absorption path $\Delta X \simeq 32.7$, in which a total of 22 lines have been detected.

In the following, we give a brief description of the new quasars and of the detected \SiIV\  absorption systems, whose parameters are given in Table~\ref{tab_linpar}. The complete analysis of the absorption systems in the new spectra will be described in a forthcoming paper (Davies et al. in prep.).  
   
\smallskip
\noindent
\begin{center}
{\it ATLAS J025.6821-33.4627 (J0142-3327)}
\end{center}

\noindent
This object was discovered in the context of the VST ATLAS survey \citep{carnall15}. It has a redshift $z_{\rm em} =6.3379$ measured from the  [C\,{\sc ii}] emission line \citep{decarli18}.   
It was observed with XSHOOTER at the VLT in October/November 2015 and in November 2018.  
We detect in the spectrum several \CIV\ absorption systems of which two, at $z$=5.64573 and $z$=5.76781, have also an associated \SiIV\ absorption. 

\smallskip
\noindent
\begin{center}
{\it VDES J0224-4711 }
\end{center}

\noindent
This quasar, at $z=6.526$, was discovered recently \citep{reed17} and is the second most luminous quasar known at $z \ge 6.5$. 
It was observed with XSHOOTER at the VLT in November 2017, January 2018 and January/December 2019. 
Several \CIV\ systems have been detected in this spectrum with that at $z$=6.03082 showing an associated \SiIV\ absorption.
We detect also a possible \SiIV\ absorption at $z$=5.55276, the corresponding \CIV\ transitions are not detected because they would fall in a very noisy region of the spectrum.  
 
\smallskip

\noindent
\begin{center}
{\it PSO J036.5078+03.0498 (J0226+0302) }
\end{center}

\noindent
\citet{venemans15} discovered this very bright quasar based on the  imaging obtained by the Pan-STARRS1 Survey. The systemic redshift,  $z=6.5412$, was determined from the  [C\,{\sc ii}] emission line \citep{decarli18}.     
Observations with XSHOOTER at the VLT were carried out in December 2017, January/November 2018 and January 2019. 
In this spectrum, two systems have both \CIV\ and \SiIV\ lines at $z$=5.89869 and $z$=5.902399. 

\begin{table}
\caption{The XSHOOTER $z\sim6$ sample. The columns are the same as in Table~\ref{tab_highres}.}
\begin{minipage}{75mm}
\label{tab_highz}
\begin{tabular}{l l l l c c}
\hline  
Object & $z_{\rm em}$ & $z_{\rm min}$ & $z_{\rm max}$ &  $R_{\rm new}$ & Ref. \\ 
\hline
SDSS J0836+0054 & 5.810 & 4.960 & 5.697 & 13100 & 1\\ 
ULAS J0148+0600 & 5.98 & 5.108 & 5.864 &  13300 & 2, 3 \\ 
SDSS J1306+0356 & 6.0337 & 5.155 & 5.917 &  12000 & 1 \\ 
SDSS J0818+1722 & 6.02 &  5.143 & 5.904 &  11000 & 1 \\ 
CFHQS J1509-1749 & 6.1225 & 5.233 & 6.005 & 11800 & 1 \\ 
ULAS J1319+0950 & 6.1330 & 5.242 & 6.015 & 13700 & 1 \\ 
SDSS J1030+0524 & 6.308 & 5.395 & 6.187 & 12300 & 1 \\ 
ATLAS J025-33 & 6.3379 & 5.422 &  6.216 & 11200 & 3\\ 
VDES J0224-4711 &  6.526 & 5.586 &  6.401 & 11200 & 3 \\ 
PSO J036+03 & 6.5412 &  5.599 &  6.416 & 10700 & 3 \\ 
\hline
\end{tabular}
1 \citet{dodorico13}; 2 \citet{codoreanu18}; 3 This work
\end{minipage}
\end{table}

\begin{table}
\caption{Log of observations for the 3 new quasars at $z>6.3$ observed with XSHOOTER.  Columns 2 and 4 report the cumulative exposure time per target, per run. While columns 3 and 5 report the adopted slit width.  They correspond to nominal resolving powers of $R \simeq 8900$ for 0.9" in the VIS, and $R \simeq 5600$ (8100) for  0.9" (0.6") in the NIR.   }
\begin{minipage}{75mm}
\label{tab_obslog}
\begin{tabular}{l c l c l}
\hline  
Program & Exp.T$_{\rm VIS}$ & slit$_{\rm VIS}$ &  Exp.T$_{\rm NIR}$ & slit$_{\rm NIR}$\\ 
& (sec) & (") & (sec) & (") \\
\hline
\multicolumn{5}{c}{{\it ATLAS J025.6821-33.4627}} \\
P096.A-0418$^a$  & 5580 & 0.9 & 5760 & 0.9JH \\
P0102.A-0154$^b$ & 14160 & 0.9 & 14400 & 0.6 \\
&&&&\\
\multicolumn{5}{c}{{\it VDES J0224-4711}} \\
P0100.A-0625$^b$  & 4640 & 0.9 & 4800 & 0.9 \\
P0102.A-0154$^b$ & 4720 & 0.9 & 4800 & 0.6 \\
P1103.A-0817$^b$ & 24000 & 0.9 & 24000 & 0.6 \\
&&&&\\
\multicolumn{5}{c}{{\it PSO J036.5078+03.0498}} \\
P0100.A-0625$^b$  & 4640 & 0.9 & 4800 & 0.9 \\
P0102.A-0154$^b$ & 18880 & 0.9 & 19200 & 0.6 \\
\hline
\end{tabular}
$^a$ P.I. Shanks; $^b$ P.I. D'Odorico V.
\end{minipage}
\end{table}

\begin{table}
\caption{Results of the \SiIV\ lines identification and fitting for the XSHOOTER spectra of J0148 and the 3 new quasars at $z>6.3$.  }
\begin{minipage}{65mm}
\label{tab_linpar}
\begin{tabular}{c c c}
\hline  
$z_{\rm abs}$ & $b$ (\kms)& $\log N($\SiIV$)$ \\
\hline
\multicolumn{3}{c}{{\it ULAS J0148+0600}}  \\
$5.12517 \pm 0.00002$ &  $27. \pm 2.$ &  $12.77 \pm 0.02$ \\
$5.48778 \pm 0.00003$ &  $24. \pm 2.$ & $12.77 \pm 0.03$ \\
\multicolumn{3}{c}{{\it ATLAS J025.6821-33.4627}}  \\
 $5.64574 \pm0.00004^a$ &  5.0 &   $12.23\pm0.10$ \\
 $5.76786\pm0.00004$ & $12\pm 4$ &   $12.72\pm 0.04$ \\
\multicolumn{3}{c}{{\it VDES J0224-4711}} \\
$5.55276\pm 0.00004^a$ & $26\pm 3$ & $12.72 \pm 0.03$ \\  
$6.03082\pm 0.00006$ &  $18\pm 5$ &  $12.95 \pm0.08$ \\
\multicolumn{3}{c}{{\it PSO J036.5078+03.0498}} \\
 $5.89873\pm 0.00006$ &  6.0  &  $12.67 \pm 0.09$ \\
 $5.90223\pm 0.00001$ & $33\pm 8$ & $12.86 \pm 0.09$ \\
\hline
\end{tabular}
$^a$ This line has not been considered in the analysis because its column density or redshift fall outside the considered ranges. 
\end{minipage}
\end{table}

\subsection{Comparison between XSHOOTER and UVES spectra}


\begin{figure}
\begin{center}
\includegraphics[width=9cm]{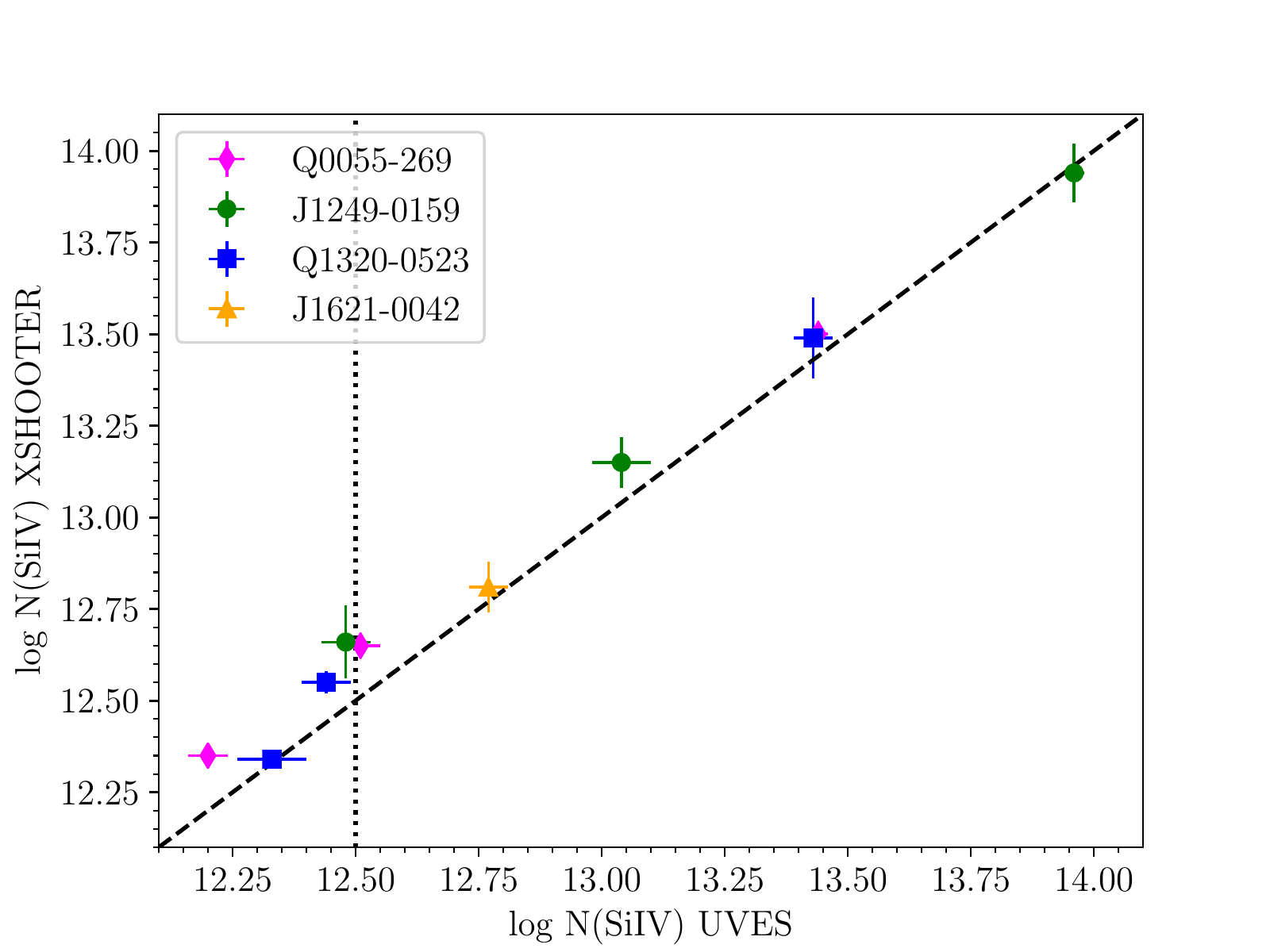}
\caption{Comparison between the column densities obtained for the \SiIV\ systems in the 4 quasars which are in common between the UVES/HIRES sample and the XQ-100 one. The vertical dotted lines marks the minimum \SiIV\ column density we adopt for our computations and the dashed line indicate the 1:1 relation.  }
\label{fig_cdcomp}
\end{center}
\end{figure}

In general, \SiIV\ absorptions are relatively weak, so even if the
resolution of XSHOOTER does not allow to resolve most of the metal
lines, our measured column densities should not be significantly
affected by unresolved saturation.  To prove this statement, we have considered the
\SiIV\ absorption lines identified and fitted in the 4 quasars that
are in common between the high-resolution sample and the XQ-100
sample: J1249-0159, Q0055-269 (or J0057-2643), J1621-0042 and J1320-0523. 
In order to compare the measured column densities, due to the
different resolutions of the two samples, we have computed the total
column density of each absorption ``system'' obtained after the merging process described in Sec. 2.1. 
Results are reported in Fig.~\ref{fig_cdcomp}. All systems with column density $\log N($\SiIV$) \gsim 12.5$ have UVES and XSHOOTER measurements which are consistent within observational errors. As we will see in the following sections, our analysis will be based only on systems with $\log N($\SiIV$) \ge 12.5$ for which we are substantially complete given the minimum SNR of the analysed spectra; as a consequence we will assume that XSHOOTER column densities are reliable even though lines are probably not resolved in XSHOOTER spectra.

\section{Statistics of the \SiIV\ absorption lines}

\begin{figure}
\begin{center}
\includegraphics[width=8.5cm]{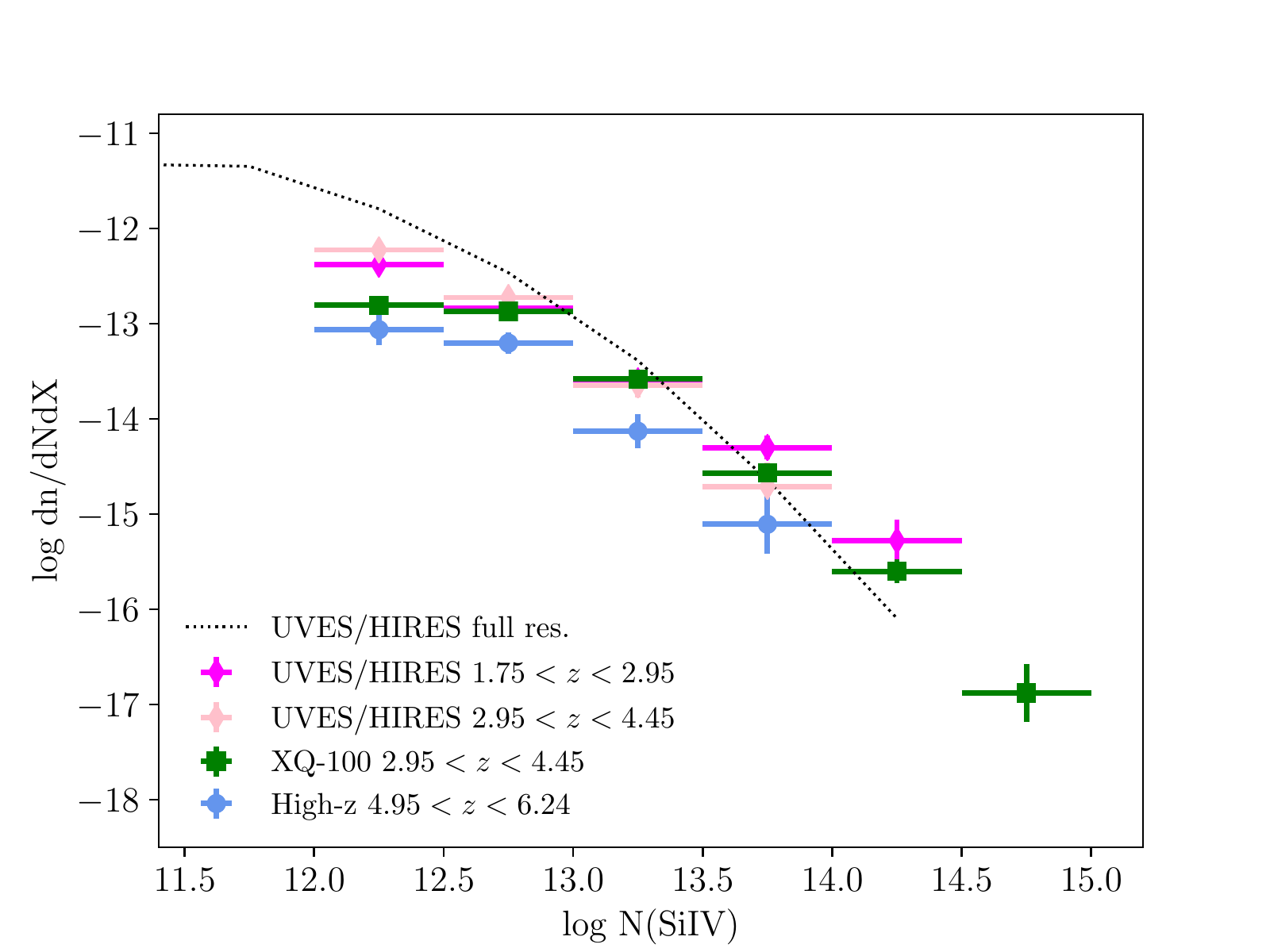}
\caption{Column density distribution function for the \SiIV\ absorptions in the UVES/HIRES (pink and magenta diamonds), XQ-100 (green quares) and $z\sim6$ XSHOOTER (blue circles) samples.The dotted line indicates the CDDF of the original UVES/HIRES sample (before merging the lines closer than 50 \kms).}
\label{fig_cddf_xq100}
\end{center}
\end{figure}


\begin{figure}
\begin{center}
\includegraphics[width=8.5cm]{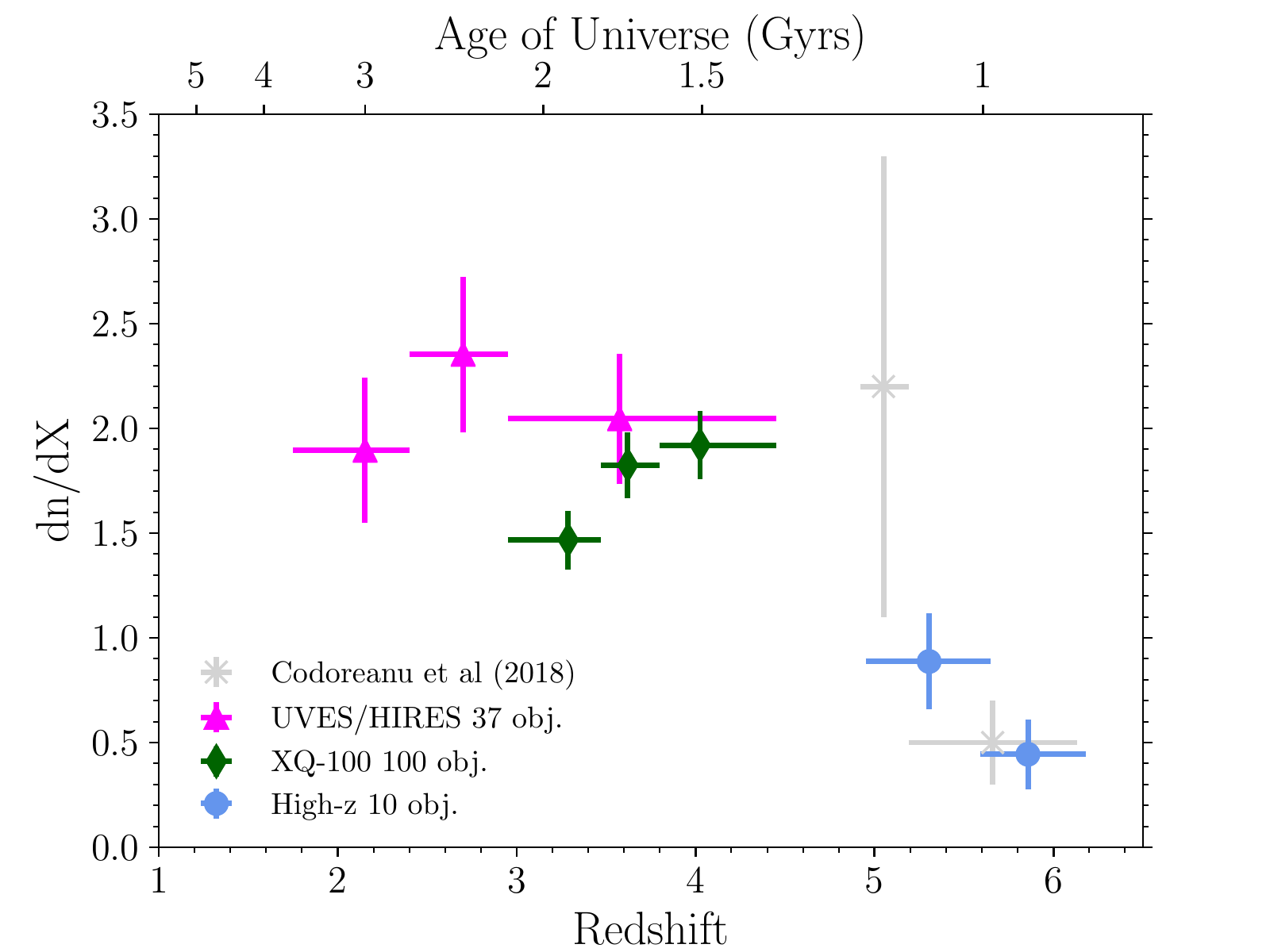}
\caption{Number density for the \SiIV\ absorptions detected in our three samples. Lines have been selected to have column densities $\log $N(\SiIV)$ \ge 12.5$ and redshifts in the range $1.75 \le z \le 6.24$. The light gray crosses are the results from \citet{codoreanu18}.} 
\label{fig_ndens}
\end{center}
\end{figure}

\subsection{Column density distribution function}

The column density distribution function (CDDF), $f(N)$, is defined as
the number of lines per unit column density and per unit redshift
absorption path, $dX$ \citep{tytler87}. 
The redshift absorption path is used to remove the redshift dependence
in the sample and put everything on a comoving coordinate scale. In the assumed
cosmology it is defined as:

\begin{equation}
\label{abs_path}
dX \equiv (1+z)^2 [ \Omega_{\rm m}(1+z)^3 + \Omega_{\Lambda}]^{-1/2}
dz.
\end{equation}  

With the adopted definition, $f(N)$ does not evolve at any redshifts
for a population whose physical size and comoving space density are
constant. 

The CDDF is a fundamental statistical property of absorption lines, similar for many aspects to the
luminosity function for stars and galaxies.

The results of the computation of the CDDF for our three samples are shown in Fig.~\ref{fig_cddf_xq100}. In the figure, the CDDF of the original UVES/HIRES sample (before merging the lines closer than 50 \kms) is shown as a dotted line. The "full resolution" sample shows an increasing CDDF, suggesting a good completeness down to the $ \log N($\SiIV$) =[11.5-12.0]$ column density bin, and then it flattens out. These values are in agreement with the column density thresholds derived from the SNR of the spectra in Section~2.1.


We observe that the number density of lines for the two UVES/HIRES redshift bins and  for XQ-100-noDLA are consistent in the column density range $12.5 \leq \log N($\SiIV$) \leq 14.5$; at $ \log N($\SiIV$) \leq 12.5$ the different behaviour is driven by the fact that the XQ-100-noDLA sample is not complete \citep[e.g.][]{ellison00}; at $ \log N($\SiIV$) \ge 14.5$ only the XQ-100-noDLA sample shows detected lines, attesting that these high column density lines are rare and a large number of lines of sight are needed to detect a significant number of them. 

Comparing the CDDF at $z<4.8$ with that at $z \ge 4.8$ computed with the $z\sim6$ XSHOOTER sample, we remark that at high $z$ the CDDF is systematically lower than at lower $z$ by $\sim 0.3-0.5$ dex for all the column density bins between 12.0 and 14.0.     The very low value at $12.0 \leq \log N($\SiIV$) \leq 12.5$ can still be explained with the incompleteness of the sample, while the lack of lines at $ \log N($\SiIV$) \ge 14.5$ is consistent with the rarity of these lines and the low number of the inspected lines of sight. 

In more detail, from the XQ-100-noDLA CDDF, a number density of $0.05\pm0.01$  ($0.009\pm0.006$, based on 2 lines) in the column density bin [14.0,14.5] ([14.5,15.0]) can be derived. Based on the surveyed absorption paths, these number densities imply a number of absorption lines consistent with zero in the column density bin [14.5,15.0] for both the $z\sim6$ XSHOOTER sample ($n=0.3\pm0.2$) and the two inspected redshift bins of the UVES/HIRES sample ($n=0.3\pm0.2$ and $0.2\pm0.1$).  In the lower column density bin [14.0,14.5], the number of expected lines is consistent with $\sim0-2$ and only in the case of the UVES/HIRES lower redshift bin, $z=[1.75,2.95]$, we actually detect one \SiIV\ absorption line.            

\subsection{Number density}

We have also computed the number density of \SiIV\ absorption lines with column densities $\log N($\SiIV$) \ge 12.5$ for which all samples have a high completeness rate. For all samples we have considered the \SiIV\ systems defined as described in Sec. 2.1.  

The number density has been computed as the number of lines ($N_{\rm lin}$) in the considered redshift bin divided by the redshift absorption path ($\Delta X$) obtained integrating eq.~\ref{abs_path} in the redshift intervals contributed by each line of sight to the specific redshift bin.   
The results are shown in Fig.~\ref{fig_ndens}  and in Table~\ref{tab_omega}. The high-resolution UVES-HIRES sample shows in the redshift range $1.75 \le z \le 4.45$ a behaviour which is consistent with a flat evolution with a possible hint of slightly decreasing number density toward lower redshifts. The XQ-100 sample which is characterized by the largest number of lines of sight and, accordingly, of absorption lines, is in agreement with the previous sample within observational errors. 
The lower value in the $z=[2.95, 3.47]$ redshift bin  could be marginally affected by the fact that it covers the first order of the XSHOOTER VIS spectrum, which in general has a slightly lower SNR with respect to the rest of the investigated spectrum. 

Considering now the $z\sim6$ sample, the measured number density values in the two analysed redshift bins are consistent between themselves within measured errors. On the other hand, $dn/dX$ at $z=[4.95,5.65]$ is  a factor of $\sim2$ lower than the XQ-100 $dn/dX$ value at $z=[3.90, 4.70]$ significant at $4.5\,\sigma$, confirming that $dn/dX$ evolved dramatically at $z>4$.

\begin{figure}
\begin{center}
\includegraphics[width=8.5cm]{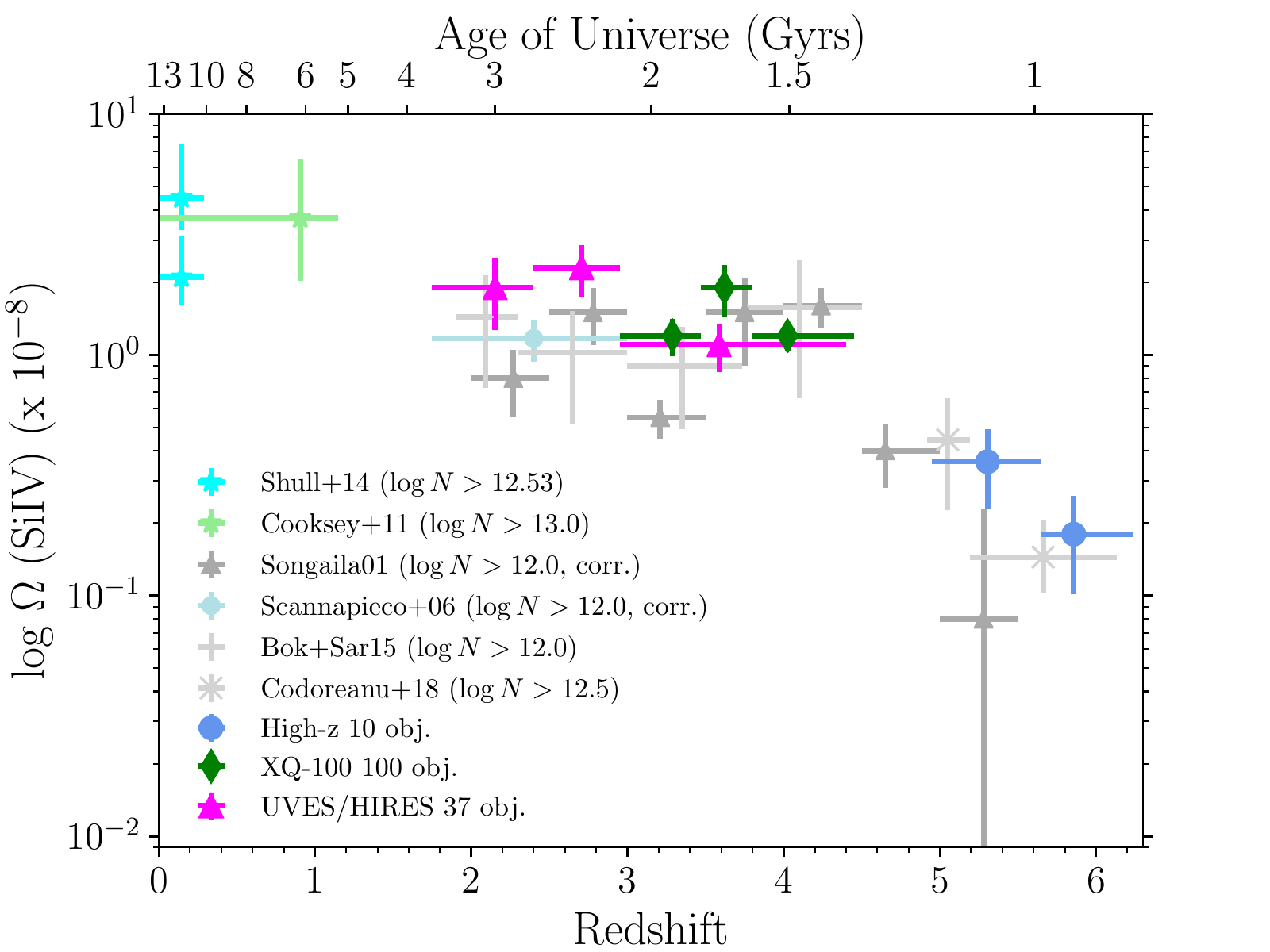}
\caption{Cosmic mass density for the \SiIV\ absorptions detected in our three samples. Lines have been selected to have column densities $\log $N(\SiIV)$ \ge 12.5$ and redshifts in the range $1.75 \le z \le 6.24$. Also shown are results  by \citet{shull14} and \citet{cooksey11} for $z \le 1$ and, for  $z \ge 1.8$, the results by \citet{songaila2001}, \citet{scannapieco}, \citet{BS15} and \citet{codoreanu18}. See the main text for further details.}
\label{fig_omega}
\end{center}
\end{figure}


\begin{table}
\caption{\SiIV\ number density and cosmic mass density results considering $\log N($\SiIV$) \ge 12.5$.}
\begin{minipage}{75mm}
\label{tab_omega}
\begin{tabular}{c l r c c}
\hline  
$z$ range & $\Delta X$ & $N_{\rm lin}$ & $dn/dX$ & $\Omega$(\SiIV)   \\ 
& & && $(\times 10^{-8})$  \\
\hline
\multicolumn{5}{c}{UVES/HIRES} \\
\hline
$1.75-2.40$ &  15.83 &  30 & $1.9\pm0.3$ & $1.9\pm0.6$ \\
$2.40-2.95$ &  17.00  & 40 & $2.3\pm0.4$ & $2.3\pm0.6$ \\
$2.95-4.40$ &  21.02  & 43 & $2.0\pm0.3$ & $1.1\pm0.2$ \\
\hline
\multicolumn{5}{c}{XQ-100 noDLA} \\
\hline
$2.95 -3.47$  & 74.96 & 110  & $1.5\pm0.1$ & $1.2\pm0.2$ \\
$3.47 -3.80$  & 73.42 & 134  & $1.8\pm0.2$ & $1.9\pm0.5$ \\
$3.80 -4.45$ & 73.43  & 141  & $1.9\pm0.2$ & $1.2\pm0.2$ \\
\hline
\multicolumn{5}{c}{XSHOOTER $z\sim6$} \\
\hline
$4.95-5.65$ &  16.91 & 15 & $0.9\pm0.2$ & $0.4\pm0.1$ \\
$5.65-6.24$ &  15.77 & 7 &  $0.4\pm0.2$ & $0.18\pm0.08$ \\
\hline
\end{tabular}
\end{minipage}
\end{table}

\section{The redshift evolution of the \SiIV\ mass density}

Finally, the CDDF can be integrated in order to
obtain the cosmological mass density of \SiIV\ in QSO absorption
systems as a fraction of the critical density today, or the contribution of \SiIV\ to the closure density:

\begin{equation}
\label{omega}
\Omega_{\rm SiIV} = \frac{H_0\, m_{\rm SiIV}}{c\, \rho_{\rm crit}} \int N
f(N) {\rm d}N, 
\end{equation}
where $H_0 = 100\, h$ \kms Mpc$^{-1}$ is the Hubble constant, $ m_{\rm
  SiIV}$ is the mass of a \SiIV\ ion, $c$ is the speed of light,
$\rho_{\rm crit} = 1.88 \times 10^{-29} h^2$ g cm$^{-3}$ and $f(N)$ is
the CDDF.  
The above integral can be approximated by the sum: 

\begin{equation}
\label{omega_approx}
\Omega_{\rm SiIV} = \frac{H_0\, m_{\rm SiIV}}{c\, \rho_{\rm crit}}
\frac{\sum_i  N_i (\mbox{\SiIV})}{\Delta X},
\end{equation}
with an associated fractional variance:

\begin{equation}
\label{omega_err}
\left( \frac{\delta \Omega_{\rm SiIV}}{\Omega_{\rm SiIV}} \right)^2 = \frac{\sum_i  
[N_i (\mbox{\SiIV})]^2}{\left[\sum_i  N_i (\mbox{\SiIV})\right]^2}
\end{equation}
as proposed by \citet{storrie96}. Note that the errors determined with this
  formula could be underestimated, in particular in the case of
  small line samples. In \citet{dodorico10}, we found that errors
  on  $\Omega_{\rm CIV}$ computed with a bootstrap technique were, at maximum, a factor of $\sim 1.5$ larger than those estimated with equation
  (\ref{omega_err}). For a fair comparison with previous results,
  however, we report in Table~\ref{tab_omega} the errors computed 
  with equation (\ref{omega_err}).

It is interesting to point out the complementarity of the information conveyed by the number density, $dn/dX$, which is heavily weighted toward abundant low-column density systems, and by the cosmological mass density, $\Omega_{\rm SiIV}$, whose value depends mainly on the rare high-column density absorption lines.  

As already observed for the CDDF and the number density of lines, Fig.~\ref{fig_omega} shows that the mass density parameter of \SiIV\ increases by a factor $\sim4-6$ moving from the redshift bin $z=[4.95, 5.65]$ to $z=[3.90, 4.70]$. In the redshift interval $2.94 \le z \le 4.70$ the evolution of $\Omega_{\rm SiIV}$ is consistent with a flat behaviour, with a possible further increase toward $z\sim1$ and the local Universe. 

The low-$z$ measurements were carried out by: \citet{cooksey11} using HST/STIS, HST/GHRS and FUSE data with \SiIV\ lines in the redshift range $z \sim 0.0-1.15$  and $\log N ($\SiIV$) \ge 13$, and by \citet{shull14} using HST/COS  spectra for \SiIV\ lines in the redshift range $z \sim 0.0-0.29$  with $\log N ($\SiIV$) \ge 12.53$. The reported values correspond to two \SiIV\ samples obtained with HST/STIS ($\Omega_{\rm SiIV} = 4.5^{+3.0}_{-1.2} \times 10^{-8}$) and with HST/COS ($\Omega_{\rm SiIV} = 2.1^{+1.0}_{-0.5} \times 10^{-8}$).  
  
\section{Our results into context}

\subsection{Comparison with previous \SiIV\ results}


The statistical quantity which is more safely comparable between different samples of absorption lines is the cosmic mass density, since it depends less on the resolution and the adopted fitting technique.

In Fig.~\ref{fig_omega}, we report our measurements and previous determinations of $\Omega_{\rm SiIV}$ at lower and comparable redshifts. Our results are in good agreement with the \citet{songaila2001} data, corrected for the different cosmology \citep[see also][based on the pixel optical depth technique]{songaila2005}.    

\citet{scannapieco} analysed a sample of 19 UVES spectra which are included in our high-resolution sample and give in their paper a value of $\Omega_{\rm SiIV} = (0.6\pm0.12) \times 10^{-8}$ at $\langle z \rangle = 2.4$, which is alarmingly lower than our result. A careful revision of their paper revealed a mistake in the computation of the normalization factor of $\Omega_{\rm SiIV}$ whose correction determines the new value $1.17\pm0.23 \times 10^{-8}$, consistent with our points.  

The cosmic mass density for \SiIV\ computed by \citet{BS15} is based on a sample of 9 high-resolution Keck/HIRES quasar spectra. We have included the quasars lines of sight of BS15 in our sample, with the aim of extending the high-redshift coverage. This is reflected in Fig.~\ref{fig_omega} by the optimal consistency between our points and those of BS15. 


\citet{codoreanu18} computed the statistical properties of \SiIV\ absorption lines approaching $z\sim6$. They considered a sample of 4 quasars all observed with XSHOOTER: ULAS J0148+0600, SDSS J1306+0356, ULAS J1319+0950 which are also part of our sample, and SDSS J0927+2001 ($z_{\rm em} = 5.79$).   They found 7 \SiIV\ systems with column density $ \log N($\SiIV$) \ge 12.5$ in the redshift range $4.92 \le z \le 6.13$,  of which 5 have $z\ge 5.19$. As we can see from Fig.~\ref{fig_ndens} and \ref{fig_omega} our results are consistent with their findings within the observational errors. 


\begin{table}
\caption{\CIV\ number density and cosmic mass density results for the high resolution sample \citep{dodorico10} and for the XSHOOTER $z\sim6$ sample \citep{dodorico13} considering column densities $\log N($\CIV$) \ge 13.0$ and adopting the same redshift bins used for \SiIV\ with the exception of the bin $z=[4.50,4.95]$ which is only covered for \CIV\ absorptions.}
\begin{minipage}{75mm}
\label{tab_omegaCIV}
\begin{tabular}{c l r c c}
\hline  
$z$ range & $\Delta X$ & $N_{\rm lin}$ & $dn/dX$ & $\Omega$(\CIV)   \\ 
& & && $(\times 10^{-8})$  \\
\hline
\multicolumn{5}{c}{UVES/HIRES} \\
\hline
$1.75-2.40$ & 32.43 & 127 & $3.9\pm0.3$ & $14\pm3$  \\
$2.40-2.95$ & 21.32 &  90 & $4.2\pm0.4$ & $14\pm3$ \\
$2.95-4.45$ & 23.98 & 102 & $4.2\pm0.4$ & $6.1\pm0.8$  \\ 
\hline
\multicolumn{5}{c}{XSHOOTER $z\sim6$ } \\
\hline
$4.50-4.95$ & 12.21 & 53 & $4.3\pm0.6$ & $2.9\pm0.5$ \\
$4.95-5.65$ & 23.73 & 41 & $1.7\pm0.3$ & $2.0\pm0.8$ \\ 
$5.65-6.24$ & 16.28  & 14 & $0.9\pm0.2$ & $0.7\pm0.3$ \\
\hline
\end{tabular}
\end{minipage}
\end{table}

\begin{figure}
\begin{center}
\includegraphics[width=9cm]{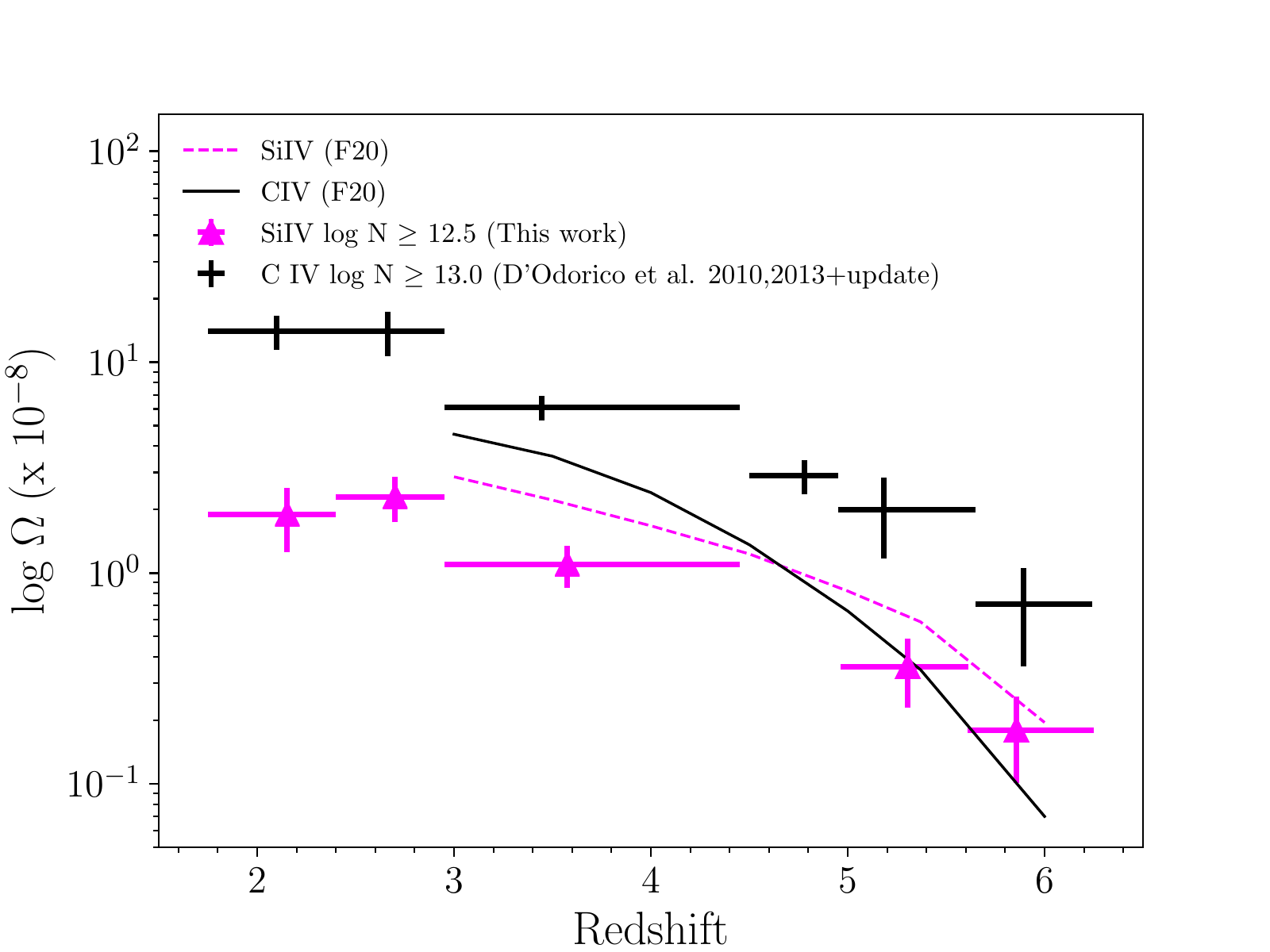}
\caption{Comparison of the cosmic mass density parameters, $\Omega$, for \SiIV\ and \CIV\ \citep[][this work]{dodorico10,dodorico13} among themselves and with the predictions of the simulations by \citet{finlator20}.   }
\label{fig_omega_SiIVCIV}
\end{center}
\end{figure}

\begin{figure}
\begin{center}
\includegraphics[width=9cm]{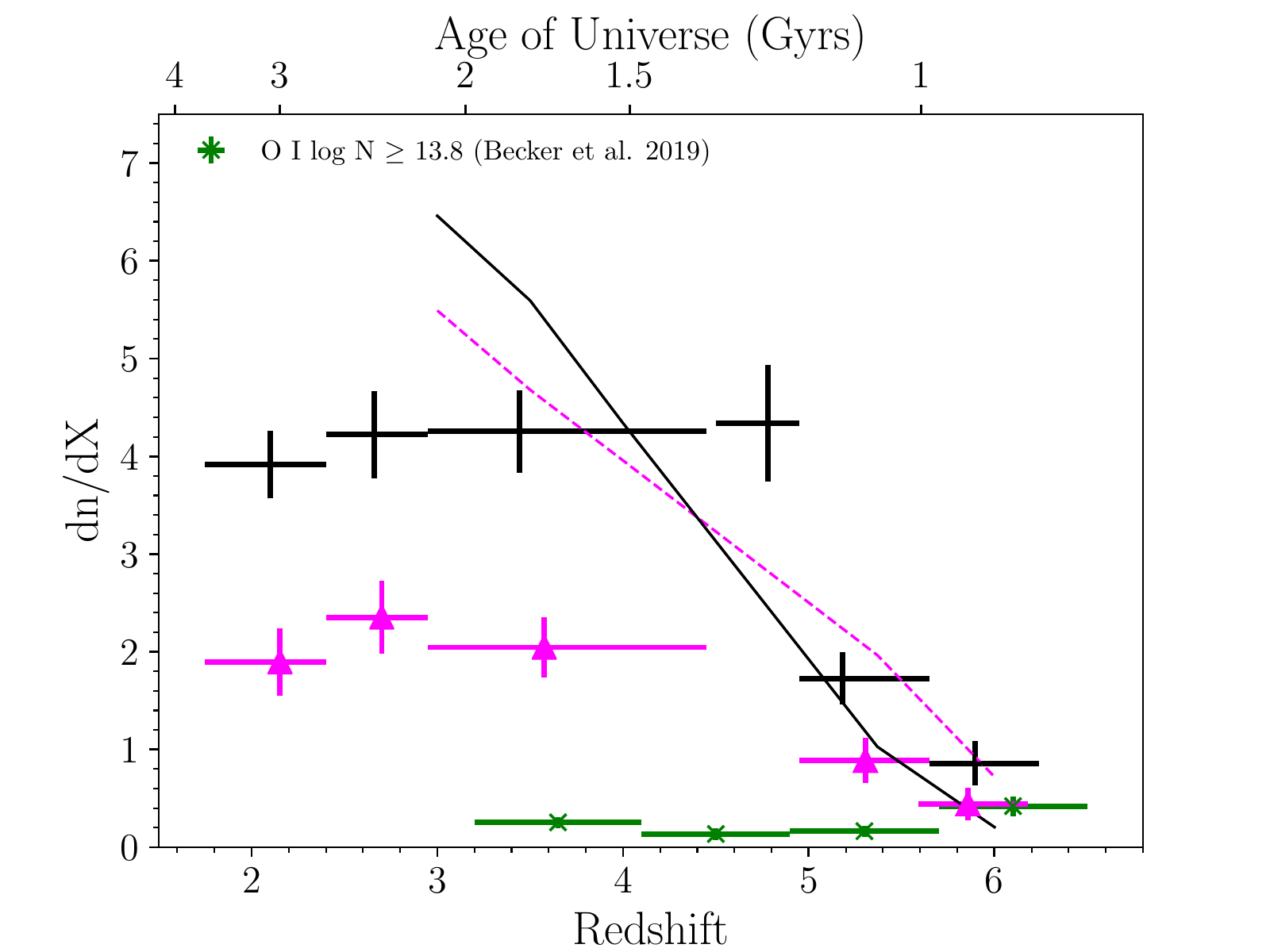}
\caption{Number densities, $dn/dX$, of \OI\ \citep[green crosses, ][]{becker19}, \SiIV\ and \CIV\ \citep[][this work]{dodorico10,dodorico13} as a function of redshift, compared with the predictions for \SiIV\ and \CIV\ by \citet{finlator20}. Symbols and lines are the same of Fig.~\ref{fig_omega_SiIVCIV}. }
\label{fig_ndens_OICIV}
\end{center}
\end{figure}

\subsection{Comparison with \CIV\ and \OI\ statistics}

The statistical properties of \CIV\ lines  were the first to be studied thanks to the large available samples.  
In \citet{dodorico13}, we investigated the evolution with redshift of the \CIV\ column density distribution function and cosmic mass density parameter up to $z\sim6$. Both quantities increase significantly between $z\sim6$ and 5 and then stay approximately constant in the range $2.0 \lsim z \lsim 5.0$. More recent studies \citep{codoreanu18,meyer19} confirm these results. 

The cosmic mass densities of \SiIV\ absorbers with $\log N \ge 12.5$ estimated in this work and of \CIV\ systems with $\log N($\CIV$) \ge 13.0$ are compared in Fig.~\ref{fig_omega_SiIVCIV}.  The choice of the \CIV\ column density threshold depends on the \SiIV\ column density threshold and on the observation that, in particular at $z\gsim 5$, the average ratio of \SiIV/\CIV\ column densities in log is $\sim -0.5$ \citep{dodorico13}.  $\Omega_{\rm CIV}$ has been computed in the same redshift bins adopted for $\Omega_{\rm SiIV}$ and it is based on the UVES/HIRES sample from \citet{dodorico10} and on the $z\sim6$ XSHOOTER sample from \citet{dodorico13}. The $z\sim6$ XSHOOTER sample has been updated with the addition of the \CIV\ lines of ULAS J0148+0600 \citep{codoreanu18} and of those of the three new quasars analysed in this work\footnote{The corresponding lists of \CIV\ absorbers are reported in the Appendix.}  The two cosmic mass densities evolve with redshift in a very similar way, with an approximately constant ratio in the whole redshift range.  



In Fig.~\ref{fig_ndens_OICIV}, we compare $dn/dX$ for \SiIV\ absorbers with $\log N \ge 12.5$, estimated in this work, with the number densities of \OI\ systems with $\log N($\OI$) \ge 13.8$ \citep{becker19} and \CIV\ absorbers with $\log N($\CIV$) \ge 13.0$. 
\SiIV\ traces the \CIV\ behaviour quite closely at all redshifts, with on average a factor $\sim2$ less lines in each redshift bin. 
The increase of $dn/dX$ with decreasing redshift for \CIV\ is generally ascribed to the combination of the increase of the ionization status and the increase of the average metallicity of the gas, as we approach the peak of star formation \citep[e.g.][]{finlator15,remetallica}. On the other hand, \OI\ shows a decrease at $z < 5.7$, then a constant behaviour to redshift 4 and a possible increase for $z < 4$. The decrease of \OI\ number density at $z < 5.7$  can only be explained by a variation in the physical properties of the gas that is transitioning from a relatively neutral state to higher ionization states~\citep{doughty2019}.   

The fact that \SiIV\ and \OI\ number densities coincide in the highest redshift bin does not imply that they trace the same absorbers, this is seen when inspecting absorption systems in our sample and also in other works \citep[e.g.][]{becker19}. A detailed analysis of the column density ratios of different ions in the same absorption systems will be carried out in a further work. It is interesting to see the sudden increase of the \CIV\ number density in the redshift bin $z=[4.5,4.95]$, which unfortunately cannot be covered by the present \SiIV\ sample.

\begin{figure*}
\begin{center}
\includegraphics[width=8.5cm]{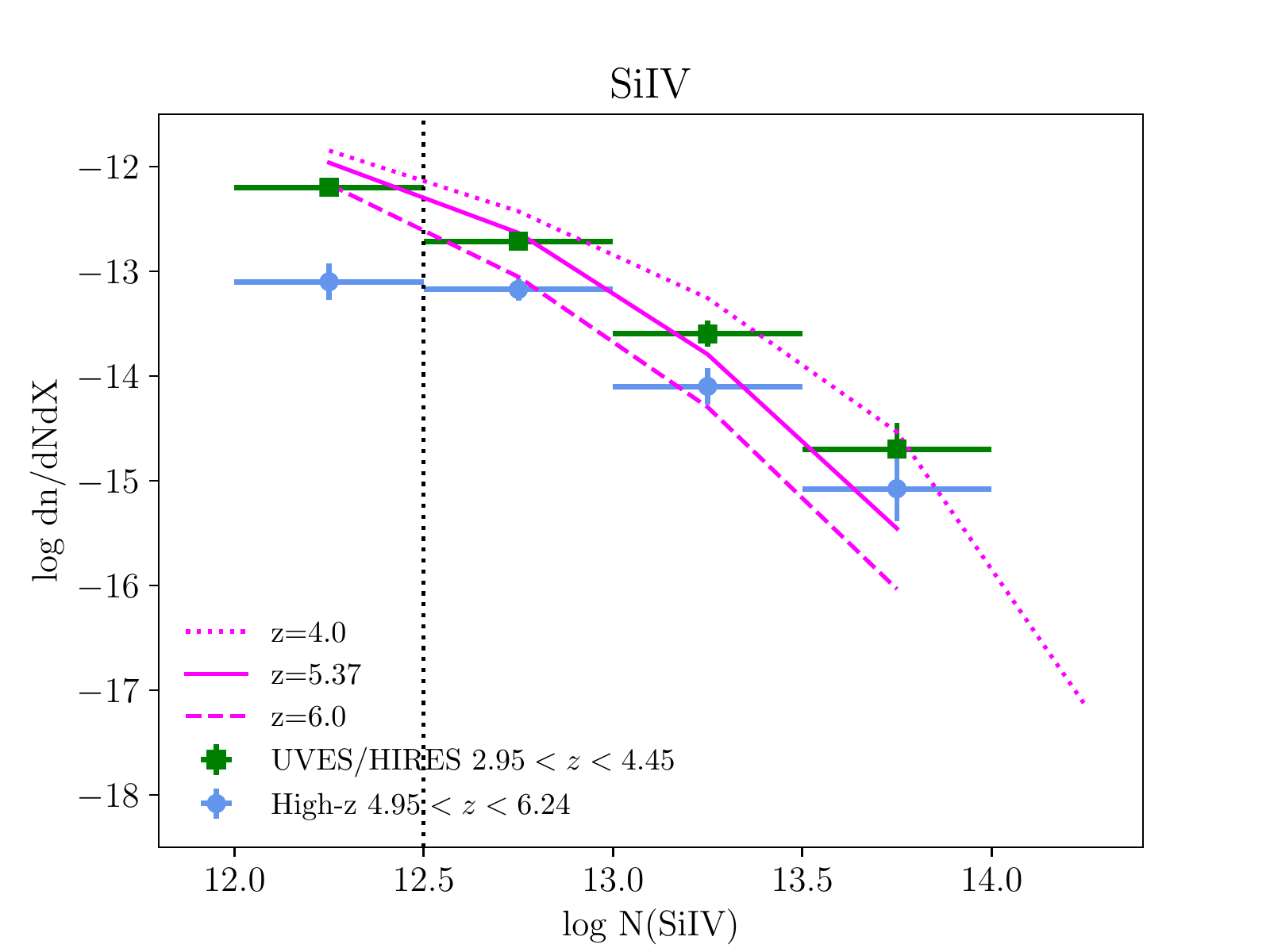}
\includegraphics[width=8.5cm]{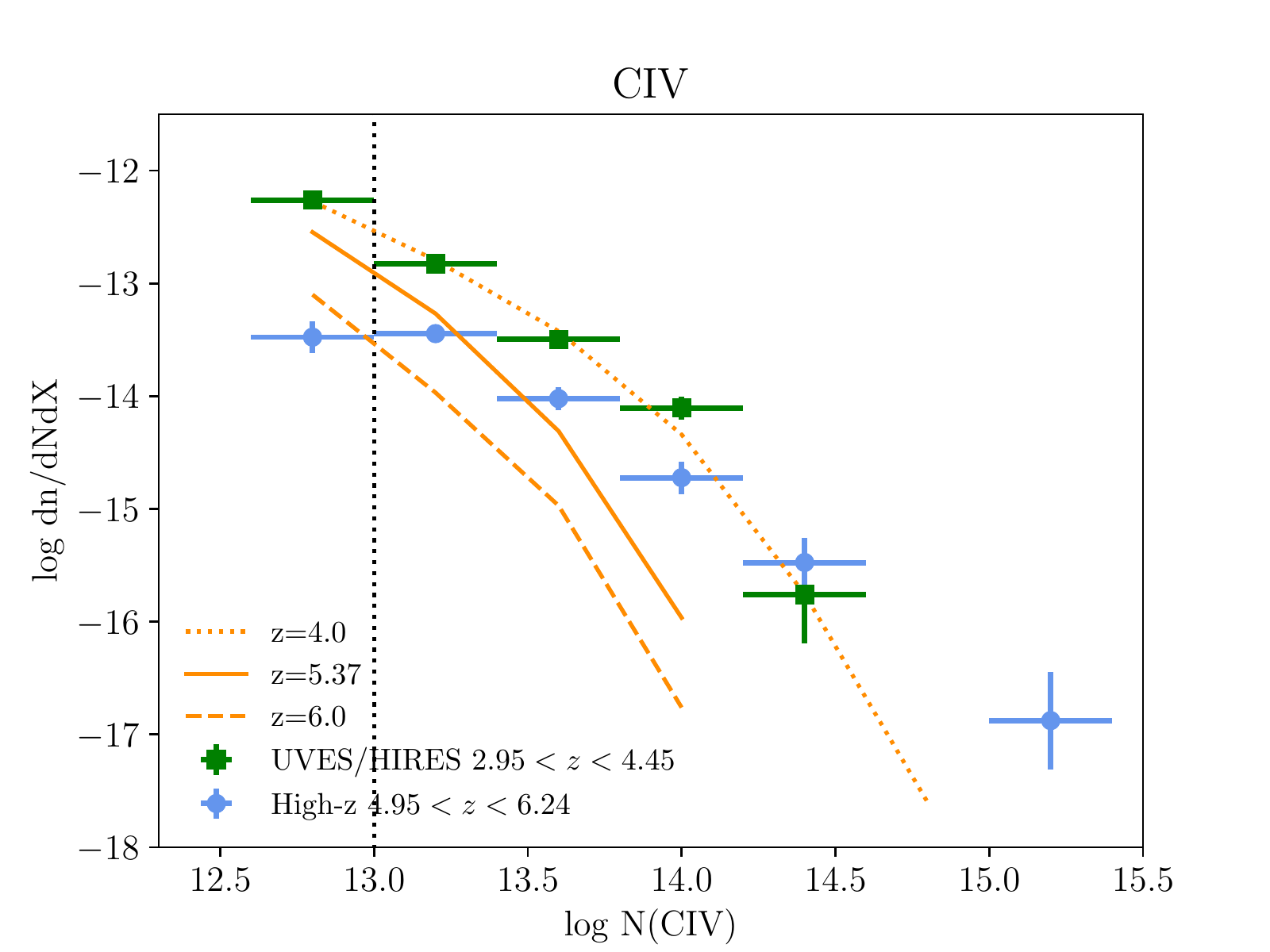}
\caption{Predictions by \citet{finlator20} for the CDDF of absorbers at three different redshifts, $z=4.0$, 5.37 and 6, compared with:  ({\it Left panel}) the CDDF of the \SiIV\ absorbers in the UVES/HIRES sample (green squares) and in the high-z sample (light blue circles) and  ({\it Right panel}) the CDDF of the \CIV\ absorbers in the high-resolution sample from \citet[][green squares]{dodorico10} and in the high-z sample \citep[][this work, light blue circles]{dodorico13}. }
\label{fig_comp_cddf}
\end{center}
\end{figure*}


\subsection{Comparison with simulation predictions}

\begin{table}
\caption{Our comparison simulations. The different columns report: the simulation reference (see text); the box size; the number of gas resolution elements when the simulation starts; the model for the ultraviolet ionizing background; the mass-loading factor; the wind velocity for a $M_b = 10^9\ M_{\odot}$ galaxy at $z=5$ and the cosmological power spectrum normalization.}
\begin{tabular}{l|cccccc}
\hline
ref & box & $N_{\mathrm{gas}}$ & UVB & $\eta_{10}$ & $v_\mathrm{wind}$ & $\sigma_8$\\
& (Mpc/$h$) & & (km/s) & & (Mpc/$h$) \\
\hline
O09   &16     & $512^3$       & HM01  & 3.96  & 82.3  & 0.83 \\
G17   &18     & $512^3$       & HM12  & 11.3  & 110.9 & 0.816 \\ 
R16   &100    & $1504^3$      & HM01  & --    & --    & 0.8288 \\ 
F20   &15     & $640^3$       & --    & 6.47  & 104   & 0.8159 \\ 
\hline
\end{tabular}
\label{table:sims}
\end{table}

We now compare our results on the evolving \SiIV\ and \CIV\ abundances against the predictions from the four cosmological hydrodynamic simulations in the literature which computed the \SiIV\ observables we have measured. 
For reference, we have summarized the simulations' relevant physical and numerical parameters in Table~\ref{table:sims}.

Large box sizes account more completely for the rare, high-column density systems that dominate $\Omega_{\rm SiIV}$ and $ \Omega_{\rm CIV}$, while high mass resolution accounts more completely for the weak systems that dominate $dn/dX$~\citep{keating16, finlator20}.
Note that an increase of the mass-loading factor\footnote{The mass-loading factor, $\eta_{10}$, quantifies the ratio of the rate at which a galaxy ejects its interstellar medium (ISM) to its star formation rate.} suppresses the star formation efficiency and the CGM metallicity, decreasing the abundance of high-ionization metal absorbers~\citep[][R16]{rahmati16}. On the other hand, the wind speed regulates the ability of galactic outflows to heat the CGM as well as the fraction of ejecta that travels to the virial radius.
The predicted CGM metallicity is proportional to the metal yield, which is uncertain by roughly a factor of two \citep{wiersma09}. 


\citet[][O09]{oppenheimer09} 
extract absorber catalogs using a homogeneous UVB~\citep[][HM01]{hm01} and a ``bubble" model in which the UVB is determined by the nearest galaxy. 
Comparing with their Figure 11, both models overpredict $\Omega_{\rm SiIV}$ at $z\sim5$ and 6 by 2-3$\sigma$. By contrast, they are in good agreement with $\Omega_{\rm CIV}$ at the same redshifts. These results could indicate that the assumed silicon yield is too high. Alternatively, they could indicate that the overall CGM metallicity is too high owing to O09's low mass-loading factor. In this case, the agreement with $\Omega_{\rm CIV}$ would indicate that the adopted HM01 UVB is weak at the high energies that regulate \CIV, cancelling the effect of the high metallicity.

The reference simulation of R16 incorporates a lower mass resolution and a larger cosmological volume than the other three models, making it less complete for weak systems and more complete for strong ones. Its feedback model enables galactic outflows to form self-consistently, hence the mass-loading factor and wind velocities are not parameterized \citep{dallavecchia12}. The predicted redshift evolution of both the \SiIV\ column density distribution function and $\Omega_{\rm SiIV}$ are in qualitative agreement with our observations. By contrast, both quantities are underpredicted for \CIV, particularly at $z > 3.5$. Our results confirm that this \CIV\ underprediction, previously noted by R16, does not reflect observational sample size limitations. In analogy with our conclusions from O09, we find that either the assumed carbon yield is too low (at $z>3.5$), or the predicted overall CGM metallicity is realistic while the HM01 UVB is, as in the case of O09, too weak at high energies.

In the fiducial model of \citet[][G17]{garcia17}, the high mass-loading factor suppresses star formation and hence the CGM metallicity. On the other hand, the high wind speeds heat the CGM efficiently. We therefore expect a partial cancellation between the impacts of low metallicity and high temperature on high-ionization CGM absorbers. 
\citet{codoreanu18} find that the \SiIV\ CDDF predicted by this model agrees with their observations in the redshift bin $z=[4.92,6.13]$ up to column densities of $\log N($\SiIV$) = 13.5$, while possibly overproducing stronger systems. The higher column densities probed by  our study enable us to confirm that the G17 \SiIV\ CDDF is too flat, overpredicting observations at column densities exceeding $\log N($\SiIV$) \sim 14$. $\Omega_{\rm SiIV}$ is consequently overpredicted by roughly an order of magnitude when integrating over columns up to $\log N($\SiIV$) = 15$. The predicted redshift evolution of $\Omega_{\rm SiIV}$ may also be more gradual than observed: G17 predict that it drops by $\approx 2\times$ from $z=4\rightarrow6$ \citep[Figure 8 of][]{codoreanu18}, whereas our observations indicate a factor $\sim 2.5-5$ decline over the same interval. Much of this decline occurs at $z>5$, where observations remain the most challenging. 

Finally, the simulation by \citet[][F20]{finlator20} combines a feedback model in which the adopted wind speeds and mass-loading factor are intermediate between G17 and O09 with a treatment for a self-consistent, spatially-inhomogeneous UVB that yields a realistic reionization history and post-reionization UVB amplitude. 

Comparisons with \citet{dodorico13} showed that, under the assumption of ionization-bounded escape \citep[see F20 and][]{zackrisson13},  the predictions by F20 underproduce strong \CIV\ absorbers while roughly reproducing strong \SiIV\ absorbers. 
The improved \SiIV\ sample presented in this work confirms the result of F20. The missing \CIV\ and \SiIV\ gas is found in lower ionization states. 
Indeed, F20 show that increasing ionization (density-bounded model) improves the agreement with the observed \CIV\ and \SiIV\ CDDF but worsen the agreement with the \CII\ absorber statistics  (their Fig.~7).

Fig.~\ref{fig_omega_SiIVCIV} shows that, at all redshifts, the F20 simulation accounts for the overall \SiIV\ mass density. 
The predicted $\Omega_{\rm SiIV}$ normalization is slightly high, but this offset falls within the range of uncertainty associated with the unknown metal yields. By contrast, the predicted $\Omega_{\rm CIV}$ evolution is steeper than observed, suggesting that the model cannot account for the observed early assembly of strong \CIV\ systems. 

Fig.~\ref{fig_comp_cddf} reveals that the predicted \SiIV\ CDDF is somewhat too steep, overproducing faint absorbers to a degree that increases with time (also seen in Fig.~\ref{fig_ndens_OICIV}). Further work will be required to determine what portion of this discrepancy reflects observational incompleteness, which manifests as a clear flattening in the observed \SiIV\ CDDF for $\log N ($\SiIV$) < 12.5$. The \CIV\ CDDF is likewise too steep. At $z >4$, this problem is compensated by its low normalization, leading to tolerable agreement with the observed $dn/dX$ (see Fig.~\ref{fig_ndens_OICIV}). At $z<4$, the overabundance of weak \CIV\ systems causes the model to overproduce $dn/dX$, in agreement with results from a complementary study that focused on the observed \CIV\ equivalent width distribution \citep{hasan20}.

\section{Summary}

In this paper, we have computed the statistical properties of a sample comprising almost 600 \SiIV\ absorption lines with column densities $\log N($\SiIV$) \ge 12.5$, detected in the spectra of 147 quasars at redshifts between $2.1 \lsim z_{\rm em} \lsim 6.5$. 

The main results of this work are the following: 
\begin{itemize}
\item The column density distribution function of \SiIV\ absorption lines does not show significant variations in the redshift range $1.7 \lsim z \lsim 4.7$, while for the highest redshift bin $4.95 \le z \le 6.24$ it is systematically lower at all column densities (Fig.~\ref{fig_cddf_xq100}).  The same behaviour was observed for \CIV\ absorption lines \citep[e.g.][]{dodorico13}.
\item The number density of lines per unit redshift absorption path, $dn/dX$, shows a small increase with redshift between $z\sim6$ and $z\sim 5.3$ and then a jump of a factor $\sim2.5$ at $z < 4.7$. Then, the number density remains approximately constant to $z\sim2$. 
\item The comparison of the number densities of \SiIV, \CIV, and \OI\ shows that \SiIV\ lines with $\log N($\SiIV$) \ge 12.5$ and \CIV\ lines with $\log N($\CIV$) \ge 13.0$ have a very similar evolution with redshift, while \OI\ shows a mild decrease for $z < 5.7$ as reported in  \citet{becker19}. 
\item The \SiIV\ cosmic mass density shows a smooth increase from redshift $\sim 6$ to $3.5$ (of a factor $\sim5$) and then it stays constant to $z\sim2$. 
\item The comparison of $\Omega_{\rm SiIV}$ with $\Omega_{\rm CIV}$ shows a similar evolution in redshift with an almost constant ratio. 

\item Finally, the examination of the predictions from cosmological, hydro-dynamical simulations indicates that the observed CDDF and cosmic mass density of \SiIV\ are well reproduced when the same quantities for \CIV\ are underpredicted. In those cases in which feedback (or other properties) are boosted to reproduce \CIV, the \SiIV\ mass density is generally overproduced.  The \CIV\ line incidence predicted by \citet{finlator20} is generally consistent with observations while the \SiIV\ line incidence is roughly consistent with observations at $z=6$ but then grows too rapidly down to at least $z=3$. Meanwhile, the predicted \CIV\ CDDF of the F20 model is too steep, particularly at early times, while the \SiIV\ CDDF has roughly the correct shape.
\end{itemize}

The results presented in this work probe new observational ground to constrain future models and advance our understanding of the distribution of metals and the enrichment mechanism of CGM and IGM. 

At $z\gsim 5$, we are still dealing with low number statistics and the results are relatively uncertain. This problem should soon be solved by the XQR-30 survey that will provide a sample of 30 new QSO spectra in the redshift range $z_{\rm em} = 5.8-6.6$  obtained with XSHOOTER at the VLT in the context of an ESO Large Programme (1103.A-0817, P.I. V. D'Odorico) whose observations are almost complete. 

In the longer run, when high-resolution spectrographs will be available at 30-40m class telescopes \citep[e.g.][]{marconiHIRES} it will be possible to carry out more detailed studies on $z\sim6$ metal absorption lines with the resolution and column density ranges reachable today with 8-10m class telescope at $z\sim2-4$. \citet{doughty18} have shown that larger ranges in column densities for \CIV\ and \SiIV\ will allow to better distinguish between different UVB models at $z\sim6$.

\section*{Acknowledgments}

We are thankful to the anonymous referee who has provided insightful comments to the paper allowing us to improve it sensibly. 
This work is based in part on observations collected at the European Southern Observatory Very Large Telescope, Cerro Paranal, Chile – Programmes Programs 65.O-0296, 166.A-0106, 069.A-0529, 079.A-0226, 084.A-0390, 084.A-0550, 085.A-0299, 086.A-0162, 087.A- 0607, 268.A-5767, 189.A-0424, 096.A-0418, 0100.A-0625, 0102.A-0154, 0103.A-0817 and 1103.A-0817.
Further observations were made at the W.M. Keck Observatory, which is operated as a scientific partnership between the California Institute of Technology and the University of California; it was made possible by the generous support of the W.M. Keck Foundation. The authors wish to recognize and acknowledge the very significant cultural role and reverence that the summit of Maunakea has always had within the indigenous Hawaiian community. 
S.L. acknowledges support by FONDECYT grant 1191232. KF  acknowledges support from the National Science Foundation (NSF) via Award Number 2006550. The Technicolor Dawn simulations were enabled by the Extreme Science and Engineering Discovery Environment, which is supported by NSF grant number ACI-1548562. The Cosmic Dawn Center is funded by the Danish National Research Foundation. 

\section*{Data Availability}
All the ESO VLT raw spectra used in this paper are publicly available from the ESO archive. The reduced UVES spectra are available in the SQUAD data base \citep{murphySQUAD}. The reduced XQ-100 spectra are available from the ESO archive, in the Phase 3 Data Releases section. The lists of fitted absorption lines underlying the obtained results, which are not published in this paper or available from the literature will be shared on reasonable request to the corresponding author.


\appendix

\section{Update on the $z\sim6$ \CIV\ absorbers}

In this section, we report the lists of \CIV\ lines with $\log N($\CIV$) \ge 13.0$, used in the present analysis from the three new spectra of quasars at $z\sim6.5$ (see Section~2.3). A more thorough analysis of all the absorbers detected in those spectra is deferred to a further paper. 

\begin{table}
\caption{New \CIV\ absorbers added to the high-$z$ sample.}
\begin{minipage}{65mm}
\label{tab_linpar}
\begin{tabular}{c c c}
\hline  
$z_{\rm abs}$ & $b$ (\kms)& $\log N($\SiIV$)$ \\
\hline
\multicolumn{3}{c}{{\it ATLAS J025.6821-33.4627}}  \\
$4.78570 \pm  0.00006$ & $ 45\pm 6$ & $13.26\pm  0.03$ \\ 
$4.86459 \pm  0.00002$ & $ 23\pm 1$ & $13.15\pm  0.02$ \\ 
$4.89616 \pm  0.00007^a$ & $-$ & $13.39\pm  0.02$ \\ 
$4.93643 \pm 0.00002^a$ & $-$ & $13.17 \pm 0.02$ \\
$5.18942 \pm 0.00004^a$ & $-$ & $13.43\pm  0.02$ \\ 
$5.20948 \pm  0.00004$ & $ 5.0 $ & $13.05\pm  0.08$ \\ 
$5.21745 \pm  0.00007$ & $ 16 \pm 6$ & $13.05\pm  0.07$ \\ 
$5.31593 \pm  0.00006$ & $ 46\pm 4$ & $13.67\pm  0.03$ \\ 
$5.31795 \pm  0.00003$ & $ 5.0 $ & $13.1\pm  0.1$ \\ 
$5.31906 \pm  0.00007$ & $ 62 \pm4$ & $13.88\pm  0.03$ \\ 
$5.64574 \pm  0.00003$ & $ 6.5 $ & $13.2\pm  0.2$ \\ 
$5.7354 \pm  0.0002$ & $ 36 \pm13$ & $13.2\pm  0.1$ \\ 
$5.76781 \pm  0.00009$ & $ 36\pm 6$ & $13.31\pm  0.05$ \\ 
$6.0996 \pm  0.0001$ & $ 39\pm 8$ & $13.34\pm  0.06$ \\  
\multicolumn{3}{c}{{\it VDES J0224-4711}} \\
$4.91166 \pm0.00003$ & $28\pm 4$  & $ 13.35\pm 0.05$ \\ 
$4.9136\pm 0.0001$ & $56 \pm20$ & $ 13.1\pm 0.1$ \\  
$4.91529\pm 0.00003$ & $19\pm 3$  & $  13.25 \pm0.04$ \\ 
$4.99276\pm 0.00005$ & $49\pm 5$  & $  13.37 \pm0.03$ \\ 
$5.00508\pm 0.00003$ &  $32 \pm4$ & $14.2\pm 0.1$ \\ 
$5.0061\pm 0.0007$ &  $58\pm 21$ &  $13.7\pm 0.3$ \\ 
$5.10894\pm 0.00003$ & $58\pm  2$ & $15.02\pm 0.04$ \\ 
$5.11325 \pm0.00003$ & $22\pm 2$ &  $14.23\pm 0.05$ \\ 
$6.03082\pm 0.00005$ & $16\pm  4$ & $13.62 \pm0.07$ \\ 
$6.17255\pm 0.00006$ & $35\pm  4$ & $13.98 \pm0.04$ \\ 
\multicolumn{3}{c}{{\it PSO J036.5078+03.0498}} \\
$4.99126\pm 0.00002$ &  $51\pm 1$ &  $14.39\pm 0.01$ \\
$5.1488 \pm0.0001$   & $61 \pm 7$  & $13.72\pm 0.05$ \\
$5.2428\pm 0.0001$ & $60\pm 7$  & $13.95\pm 0.04$ \\
$5.24474\pm 0.00006$ &  $22\pm5$  & $13.58\pm 0.07$ \\
$5.6885\pm 0.0002$ & $ 21 \pm 13$ & $13.2 \pm 0.1$ \\ 
$5.8121\pm 0.0001$ & $ 43 \pm 9$ & $13.26 \pm 0.07$ \\ 
$5.8262\pm 0.0001$   & $50\pm 8.$  & $13.44\pm 0.05$ \\
$5.89869\pm 0.00004$ &  $9\pm4$  & $13.7\pm 0.1$ \\
$5.90223\pm 0.00001$ & $ 32 \pm 8$ & $13.2 \pm 0.1$ \\ 
\hline
\end{tabular}
$^a$ Combination of two lines closer than $\Delta v = 50$ \kms. See Section~2.1.
\end{minipage}
\end{table}


\label{lastpage}

\begin{thebibliography}{99}

\bibitem[\protect\citeauthoryear{Adelberger et al.}{2005}]{adelberger}
Adelberger K. L., Shapley A. E., Steidel C. C., Pettini M., Erb D. K.,
Reddy N.  A., 2005, ApJ, 629, 636






\bibitem[\protect\citeauthoryear{Barai et al.}{2013}]{barai13} 
Barai P., Viel M., Borgani S., Tescari E., Tornatore L., Dolag K., Killedar M., et al., 2013, MNRAS, 430, 3213

\bibitem[\protect\citeauthoryear{Becker et al.}{2019}]{becker19}
Becker G. D. et al. 2019, ApJ, 883, 163





\bibitem[\protect\citeauthoryear{Berg et al.}{2021}]{berg21} 
Berg T.~A.~M., Fumagalli M., D'Odorico V., Ellison S.~L., L{\'o}pez S., Becker G.~D., Christensen L., et al., 2021, MNRAS, 502, 4009 

\bibitem[\protect\citeauthoryear{Berg et al.}{2016}]{berg16}
Berg T.~A.~M., Ellison S.~L., S{\'a}nchez-Ram{\'\i}rez R., Prochaska J.~X., Lopez S., D'Odorico V., Becker G., et al., 2016, MNRAS, 463, 3021 


\bibitem[\protect\citeauthoryear{Bergeron et al.}{2004}]{bergeron04}
Bergeron J., et al., 2004, ESO The Messenger, 118, 40 

\bibitem[\protect\citeauthoryear{Bernstein, Burles, \& Prochaska}{2015}]{Bernstein15} 
Bernstein R.~M., Burles S.~M., Prochaska J.~X., 2015, PASP, 127, 911. 

 

\bibitem[\protect\citeauthoryear{Birnboim \& Dekel}{2003}]{birnboim03} 
Birnboim Y., Dekel A., 2003, MNRAS, 345, 349

\bibitem[\protect\citeauthoryear{Boksenberg \& Sargent}{2015}]{BS15}
Boksenberg A., Sargent W. L. W, 2015, ApJS, 218, 7





\bibitem[\protect\citeauthoryear{Bordoloi et al.}{2014}]{bordoloi14}
Bordoloi R., Tumlinson J., Werk J. K., Oppenheimer B. D., Peeples M. S., Prochaska J. X., Tripp T. M., Katz N., et al.,  2014  ApJ, 796, 136





\bibitem[\protect\citeauthoryear{Calura et al.}{2012}]{calura12}
Calura F., Tescari E., D'Odorico V., Viel M., Cristiani S., Kim
T.-S., Bolton J. S., 2012, MNRAS, 422, 3019

\bibitem[\protect\citeauthoryear{Calura \& Matteucci}{2006}]{calura06} 
Calura F., Matteucci F., 2006, MNRAS, 369, 465 

\bibitem[\protect\citeauthoryear{Carnall et al.}{2015}]{carnall15}
Carnall A. C., et al. 2015, MNRAS, 451, L16
   

\bibitem[\protect\citeauthoryear{Carswell et al.}{2014}]{vpfit}
Carswell R. F., Webb J. K., 2014, VPFIT, Astrophysics Source Code Library, record ascl:1408.015

\bibitem[\protect\citeauthoryear{Cen \& Chisari}{2011}]{cen_chisari11}
Cen R. \& Chisari N. E., 2011, ApJ, 731, 11




\bibitem[\protect\citeauthoryear{Codoreanu et al.}{2018}]{codoreanu18}
Codoreanu A.,Ryan-Weber E. V., Garc\'\i a L. A.,  Crighton N. H. M., Becker G. D., Pettini M., Madau P., Venemans B., 2018, MNRAS, 481, 4940



\bibitem[\protect\citeauthoryear{Cooksey et al.}{2013}]{cooksey13}
Cooksey K. L., Kao M. M., Simcoe R. A., O'Meara J. M., Prochaska
J. X., 2013, ApJ, 763, 37

\bibitem[\protect\citeauthoryear{Cooksey et al.}{2011}]{cooksey11}
Cooksey K. L., Prochaska J. X., Thom C., Chen H.-W. 2011, ApJ 729, 87

\bibitem[\protect\citeauthoryear{Cooksey et al.}{2010}]{cooksey10}
Cooksey K. L., Thom C., Prochaska J. X., Chen H-W., 2010, ApJ, 708, 868 

\bibitem[\protect\citeauthoryear{Cooper et al.}{2019}]{cooper19}
Cooper T. J., Simcoe R. A., Cooksey K. L., Bordoloi R., Miller D. R., Furesz G., Turner M. L., Ba\~{n}ados E., 2019, ApJ, 882, 77


\bibitem[\protect\citeauthoryear{Cowie et al.}{1995}]{cowie95}
Cowie L.L., Songaila A., Kim T.-S., Hu E.M., 1995, AJ, 109, 1522


 
\bibitem[\protect\citeauthoryear{Cupani et al.}{2020}]{cupani20} 
Cupani G., D'Odorico V., Cristiani S., Russo S.~A., Calderone G., Taffoni G., 2020, SPIE, 11452, 114521U



\bibitem[\protect\citeauthoryear{Dalla Vecchia \& Schaye}{2012}]{dallavecchia12} 
Dalla Vecchia C., Schaye J., 2012, MNRAS, 426, 140 

\bibitem[\protect\citeauthoryear{Decarli et al.}{2018}]{decarli18}
Decarli R., et al., 2018, ApJ, 854, 97
 

\bibitem[\protect\citeauthoryear{Dekker et al.}{2000}]{dekker}
Dekker H., D'Odorico S., Kaufer A., Delabre B., Kotzlowski H., 2000,
SPIE, 4008, 534 

\bibitem[\protect\citeauthoryear{D'Odorico et al.}{2016}]{dodorico16}
D'Odorico V. et al., 2016, MNRAS, 463, 2690

\bibitem[\protect\citeauthoryear{D'Odorico et al.}{2013}]{dodorico13}
D'Odorico V. et al., 2013, MNRAS, 435, 1198
  
\bibitem[\protect\citeauthoryear{D'Odorico et al.}{2010}]{dodorico10}
D'Odorico V., Calura F., Cristiani S., Viel M., 2010,  MNRAS, 401, 2715


\bibitem[\protect\citeauthoryear{Doughty et al.}{2018}]{doughty18} Doughty C., Finlator K., Oppenheimer B.~D., Dav{\'e} R., Zackrisson E., 2018, MNRAS, 475, 4717

\bibitem[\protect\citeauthoryear{Doughty \& Finlator}{2019}]{doughty2019} 
Doughty, C. \& Finlator, K., 2019, MNRAS, 489, 2755 

\bibitem[\protect\citeauthoryear{Ellison et al.}{2000}]{ellison00}
Ellison S. L., Songaila A., Schaye J., Pettini M., 2000, AJ, 120, 1175




\bibitem[\protect\citeauthoryear{Faucher-Gigu{\`e}re, Kere{\v{s}}, \& Ma}{2011}]{fauchergiguere11} 
Faucher-Gigu{\`e}re C.-A., Kere{\v{s}} D., Ma C.-P., 2011, MNRAS, 417, 2982 



\bibitem[\protect\citeauthoryear{Finlator et al.}{2020}]{finlator20} 
Finlator K., Doughty C., Cai Z., D{\'\i}az G., 2020, MNRAS, 493, 3223 

\bibitem[\protect\citeauthoryear{Finlator et al.}{2018}]{finlator18}  
Finlator K., Keating L., Oppenheimer B. D., Dav\'e R., Zackrisson E., 2018, MNRAS, 480, 2628

\bibitem[\protect\citeauthoryear{Finlator et al.}{2015}]{finlator15} 
Finlator K., Thompson R., Huang S., Dav{\'e} R., Zackrisson E., Oppenheimer B.~D., 2015, MNRAS, 447, 2526 

\bibitem[\protect\citeauthoryear{Fontana \& Ballester}{1995}]{font:ball}
Fontana A., Ballester P., 1995, ESO The Messenger, 80, 37

\bibitem[\protect\citeauthoryear{Fossati et al.}{2021}]{fossati21} 
Fossati M., Fumagalli M., Lofthouse E.~K., Dutta R., Cantalupo S., Arrigoni Battaia F., Fynbo J.~P.~U., et al., 2021, MNRAS, 503, 3044 

\bibitem[\protect\citeauthoryear{Fossati et al.}{2019}]{fossati19}
Fossati M., et al. 2019, MNRAS, 490, 1451

\bibitem[\protect\citeauthoryear{Garc\`{i}a et al.}{2017}]{garcia17}
Garc\`{i}a L. A., Tescari E., Ryan-Weber E. V., Wyithe J.S.B., MNRAS, 470, 2494


%


\bibitem[\protect\citeauthoryear{Haardt \& Madau}{2012}]{hm12} 
Haardt F., Madau P., 2012, ApJ, 746, 125 

\bibitem[\protect\citeauthoryear{Haardt \& Madau}{2001}]{hm01}
Haardt F., Madau P., 2001 in ``Clusters of galaxies and the high
resdhift Universe observed in X-rays: recent results of XMM-Newton
and Chandra'', XXXVIth Rencontres de Moriond, eds. D.M. Neumann \&
J.T.T. Van



\bibitem[\protect\citeauthoryear{Hasan et al.}{2020}]{hasan20} 
Hasan F., Churchill C.~W., Stemock B., Mathes N.~L., Nielsen N.~M., Finlator K., Doughty C., et al., 2020, ApJ, 904, 44 



\bibitem[\protect\citeauthoryear{Keating et al.}{2016}]{keating16}
Keating L. C., Puchwein E., Haehnelt M. G., Bird S., Bolton J. S., 2016, MNRAS, 461, 606


\bibitem[\protect\citeauthoryear{Kere{\v{s}} et al.}{2005}]{keres05} 
Kere{\v{s}} D., Katz N., Weinberg D.~H., Dav{\'e} R., 2005, MNRAS, 363, 2 


\bibitem[\protect\citeauthoryear{Johnson et al.}{2017}]{johnson17}
Johnson S. D., Chen H.-W., Mulchaey J. S., Schaye, J., Straka L. A. 2017, ApJL, 850, L10

\bibitem[\protect\citeauthoryear{Liang \& Chen}{2014}]{liang14}
Liang C. J., Chen, H-W., 2014, MNRAS, 445, 2061

\bibitem[\protect\citeauthoryear{Lofthouse et al.}{2020}]{lofthouse19} 
Lofthouse E.~K., Fumagalli M., Fossati M., O'Meara J.~M., Murphy M.~T., Christensen L., Prochaska J.~X., et al., 2020, MNRAS, 491, 2057

\bibitem[\protect\citeauthoryear{L\'opez et al.}{2020}]{lopez19} 
L\'opez S., Tejos N., Barrientos L.~F., Ledoux C., Sharon K., Katsianis A., Florian M.~K., et al., 2020, MNRAS, 491, 4442


\bibitem[\protect\citeauthoryear{L\'opez et al.}{2018}]{lopez18}
L\'opez S. et al., 2018, Nature, 554, 493

\bibitem[\protect\citeauthoryear{L\'opez et al.}{2016}]{lopez16}
L\'opez S. et al., 2016, A\&A, 594, 91
      


\bibitem[\protect\citeauthoryear{Madau, Ferrara, \& Rees}{2001}]{madau01} 
Madau P., Ferrara A., Rees M.~J., 2001, ApJ, 555, 92. doi:10.1086/321474



\bibitem[\protect\citeauthoryear{Maiolino \& Mannucci}{2019}]{remetallica} 
Maiolino R., Mannucci F., 2019, A\&ARv, 27, 3 

\bibitem[\protect\citeauthoryear{Marconi et al.}{2021}]{marconiHIRES} 
Marconi A., Abreu M., Adibekyan V., Aliverti M., Allende Prieto C., Amado P., Amate M., et al., 2021, Msngr, 182, 27




\bibitem[\protect\citeauthoryear{Meyer et al.}{2019}]{meyer19}
Meyer R.A., Bosman S. E. I., Kakiichi K., Ellis R. S., 2019, MNRAS, 483, 19,  

\bibitem[\protect\citeauthoryear{Mongardi et al.}{2018}]{mongardi18}
Mongardi C., Viel M., D'Odorico V.,  Kim T.-S., Barai P., Murante G., Monaco P., 2018, MNRAS, 478, 3266 

\bibitem[\protect\citeauthoryear{Muratov et al.}{2017}]{muratov17} 
Muratov A.~L., Kere{\v{s}} D., Faucher-Gigu{\`e}re C.-A., Hopkins P.~F., Ma X., Angl{\'e}s-Alc{\'a}zar D., Chan T.~K., et al., 2017, MNRAS, 468, 4170 

\bibitem[\protect\citeauthoryear{Murphy et al.}{2019}]{murphySQUAD} 
Murphy M.~T., Kacprzak G.~G., Savorgnan G.~A.~D., Carswell R.~F., 2019, MNRAS, 482, 3458

\bibitem[\protect\citeauthoryear{Nelson et al.}{2013}]{nelson13} 
Nelson D., Vogelsberger M., Genel S., Sijacki D., Kere{\v{s}} D., Springel V., Hernquist L., 2013, MNRAS, 429, 3353 

\bibitem[\protect\citeauthoryear{Oppenheimer, Dav\'e \& Finlator}{2009}]{oppenheimer09}
Oppenheimer B.D., Dav\'e R., Finlator K., 2009, MNRAS, 396, 729


\bibitem[\protect\citeauthoryear{Oppenheimer \& Dav\'e}{2006}]{opp_dave06}
Oppenheimer B.D., Dav\'e R., 2006, MNRAS, 373, 1265



\bibitem[\protect\citeauthoryear{P{\'e}roux \& Howk}{2020}]{peroux_araa20} 
P{\'e}roux C., Howk J.~C., 2020, ARA\&A, 58, 363 

\bibitem[\protect\citeauthoryear{Perrotta et al.}{2016}]{perrotta16} 
Perrotta S., D'Odorico V., Prochaska J.~X., Cristiani S., Cupani G., Ellison S., L{\'o}pez S., et al., 2016, MNRAS, 462, 3285 







\bibitem[\protect\citeauthoryear{Prochaska et al}{2011}]{prochaska11}
Prochaska J. X., Weiner B., Chen H.-W., Mulchaey J., Cooksey K., 2011, ApJ, 740, 91

\bibitem[\protect\citeauthoryear{Rahmati et al.}{2016}]{rahmati16}
Rahmati A., Schaye J.,  Crain R. A., Oppenheimer B. D., Schaller M., Theuns T., 2016, MNRAS, 459, 310




\bibitem[\protect\citeauthoryear{Reed et al.}{2017}]{reed17}
Reed S. L. et al. 2017, MNRAS, 468, 4702




\bibitem[\protect\citeauthoryear{Rudie et al.}{2019}]{rudie19}
Rudie G. C., Steidel C. C., et al. 2019,  ApJ, 885, 61




\bibitem[\protect\citeauthoryear{Ryan-Weber et al.}{2009}]{ryanweber09}
Ryan-Weber E.V., Pettini M., Madau P., Zych B.J., 2009, MNRAS, 395, 1476


\bibitem[\protect\citeauthoryear{S\'anchez-Ram\'\i rez et al.}{2016}]{sanchez16}
S\'anchez-Ram\'\i rez R., et al. 2016 MNRAS 456, 4488

\bibitem[\protect\citeauthoryear{Savage \& Sembach}{1991}]{AOD} 
Savage B.~D., Sembach K.~R., 1991, ApJ, 379, 245


\bibitem[\protect\citeauthoryear{Scannapieco et  al.}{2006}]{scannapieco}
Scannapieco E., et al., 2006, MNRAS, 365, 615 


\bibitem[\protect\citeauthoryear{Schaye et al.}{2003}]{schaye03}
Schaye J., Aguirre A., Kim T-S., Theuns T., Rauch M., Sargent W.L.W.,
2003, ApJ, 596, 768





\bibitem[\protect\citeauthoryear{Shull et al.}{2014}]{shull14}
Shull J. M., Danforth C. W., Tilton E. M. 2014, ApJ, 796, 49

\bibitem[\protect\citeauthoryear{Simcoe et al.}{2011}]{simcoe11}
Simcoe R. A. et al., 2011, ApJ, 743, 21

 



\bibitem[\protect\citeauthoryear{Songaila}{2005}]{songaila2005}
Songaila A., 2005, AJ, 130, 1996

\bibitem[\protect\citeauthoryear{Songaila}{2001}]{songaila2001}
Songaila A., 2001, ApJ, 561, L153





 
  
\bibitem[\protect\citeauthoryear{Steidel et al.}{2010}]{steidel10}
Steidel C. C., et al., 2010, ApJ, 717, 289

\bibitem[\protect\citeauthoryear{Storrie-Lombardi et al.}{1996}]{storrie96}
Storrie-Lombardi L., McMahon R.G., Irwin M., 1996, MNRAS, 283, 79
 
\bibitem[\protect\citeauthoryear{Suresh et al.}{2015}]{suresh15} 
Suresh J., Bird S., Vogelsberger M., Genel S., Torrey P., Sijacki D., Springel V., et al., 2015, MNRAS, 448, 895

 
\bibitem[\protect\citeauthoryear{Tescari et al.}{2011}]{tescari11}
Tescari E., Viel M., D'Odorico V., Cristiani S., Calura F., Borgani S., Tornatore L., 2011, MNRAS, 411, 826
 
\bibitem[\protect\citeauthoryear{Theuns}{2021}]{theuns21} 
Theuns T., 2021, MNRAS, 500, 2741 



\bibitem[\protect\citeauthoryear{Tumlinson et al.}{2011}]{tumlinson11}
Tumlinson J., Thom C., Werk J. K., Prochaska J. X., Tripp T. M., Weinberg D. H., Peeples M. S., O'Meara J. M., et al.,  2011, Science, 334, 948

\bibitem[\protect\citeauthoryear{Turner et al.}{2016}]{turner16} 
Turner M.~L., Schaye J., Crain R.~A., Theuns T., Wendt M., 2016, MNRAS, 462, 2440 

\bibitem[\protect\citeauthoryear{Turner et al.}{2014}]{turner14}
Turner M. L., Schaye J., Steidel C. C., Rudie G. C., Strom A. L., 2014, MNRAS, 445, 794 

\bibitem[\protect\citeauthoryear{Tytler et al.}{1995}]{tytler95}
Tytler D., Fan X.-M., Burles S., Cottrell L., Davis C., Kirkman D.,
Zuo L., 1995, in QSO Absorption Lines, ed. G.Meylan (Springer-Verlag), 289

\bibitem[\protect\citeauthoryear{Tytler}{1987}]{tytler87}
Tytler D., 1987, ApJ, 321, 49

\bibitem[\protect\citeauthoryear{van de Voort et al.}{2011}]{vandevoort11} 
van de Voort F., Schaye J., Booth C.~M., Haas M.~R., Dalla Vecchia C., 2011, MNRAS, 414, 2458 

\bibitem[\protect\citeauthoryear{van de Voort et al.}{2012}]{vandevoort12} 
van de Voort, F., Schaye, J., Altay, G., et al.\ 2012, MNRAS, 421, 2809 





\bibitem[\protect\citeauthoryear{Venemans et al.}{2015}]{venemans15}
Venemans B. P. et al., 2015, ApJ, 801, L11






\bibitem[\protect\citeauthoryear{Werk et al.}{2014}]{werk14}
Werk J. K., Prochaska J. X., Tumlinson J., et al. 2014, ApJ, 792, 8

\bibitem[\protect\citeauthoryear{Wiersma et al.}{2009}]{wiersma09} 
Wiersma R.~P.~C., Schaye J., Theuns T., Dalla Vecchia C., Tornatore L., 2009, MNRAS, 399, 574 


\bibitem[\protect\citeauthoryear{Zackrisson, Inoue, \& Jensen}{2013}]{zackrisson13} 
Zackrisson E., Inoue A.~K., Jensen H., 2013, ApJ, 777, 39 
\end{thebibliography}
\end{document}